\documentclass[11pt,a4paper]{article}

\pdfoutput=1
\usepackage{color}
\usepackage{amsmath,amssymb,bm,bold-extra}
\usepackage{slashed,relsize}
\usepackage{epsfig}
\usepackage{multirow}
\usepackage{blkarray}
\usepackage{booktabs}
\usepackage{makecell}
\usepackage{cite}
\usepackage{textcomp} 

\definecolor{myblue}{rgb}{0, 0, 0.7}
\usepackage[colorlinks=true,linkcolor=black,citecolor=black,urlcolor=myblue]{hyperref}

\allowdisplaybreaks

\def\beq{\begin{equation}\displaystyle}
\def\eeq{\end{equation}}
\def\bea{\begin{eqnarray}\displaystyle}
\def\eea{\end{eqnarray}}

\newlength{\bibitemsep}
\newlength{\bibparskip}
\let\oldthebibliography\thebibliography
\renewcommand\thebibliography[1]{%
  \oldthebibliography{#1}%
  \setlength{\parskip}{\bibitemsep}%
  \setlength{\itemsep}{\bibparskip}%
}

\title{Parametrized classifiers for optimal EFT sensitivity}

\date{}

\author{
Siyu Chen$^{a}$, Alfredo Glioti$^{a}$, Giuliano Panico$^{b,c}$, Andrea Wulzer$^{a,d,e}$ \\ \\
{\small\emph{$^a$ Theoretical Particle Physics Laboratory (LPTP), Institute of Physics,}}\\
{\small\emph{EPFL, Lausanne, Switzerland}}\\
{\small\emph{$^b$ INFN, Sezione di Firenze, Via G. Sansone 1, 50019 Sesto Fiorentino, Italy}}\\
{\small\emph{$^c$ Dipartimento di Fisica e Astronomia Universit\`a di Firenze,}}\\
{\small\emph{Via G. Sansone 1, 50019 Sesto Fiorentino, Italy}}\\
{\small\emph{$^d$ CERN, 1211 Geneva 23, Switzerland}}\\
{\small\emph{$^e$ Dipartimento di Fisica e Astronomia, Universit\`a di Padova, Italy}}\\
}

\begin{document}
\baselineskip=14pt
\begin{flushright}
CERN-TH-2020-122
\end{flushright}
\vspace{2em}
{\let\newpage\relax\maketitle}
\begin{abstract}
We study unbinned multivariate analysis techniques, based on Statistical Learning, for indirect new physics searches at the LHC in the Effective Field Theory framework. We focus in particular on high-energy ZW production with  fully leptonic decays, modeled at different degrees of refinement up to NLO in QCD. We show that a considerable gain in sensitivity is possible compared with current projections based on binned analyses. As expected, the gain is particularly significant for those operators that display a complex pattern of interference with the Standard Model amplitude. The most effective method is found to be the ``Quadratic Classifier'' approach, an improvement of the standard Statistical Learning classifier where the quadratic dependence of the differential cross section on the EFT Wilson coefficients is built-in and incorporated in the loss function. We argue that the Quadratic Classifier performances are nearly statistically optimal, based on a rigorous notion of optimality that we can establish for an approximate analytic description of the ZW process.
\end{abstract}

\newpage 

\begingroup
\tableofcontents
\endgroup 


\newpage
\section{Introduction}

The amazing richness of LHC data makes searching for new physics an extremely complex process.
Three main steps can be identified, taking however into account that they are strongly interconnected and not necessarily sequential in time.
First, we need a target new physics theory. In our case this is provided by the Standard Model (SM) itself, supplemented by
operators of energy dimension $d>4$ that encapsulate the indirect effects of heavy new
particles and interactions. This setup is often dubbed the SM Effective Field Theory (EFT) in the context of high-energy physics (see e.g.~Refs.~\cite{Buchmuller:1985jz,Giudice:2007fh,Grzadkowski:2010es}). However the EFT approach is extremely common and widely employed in many other domains, eminently in Flavor Physics. The methodologies discussed in this paper could thus find applications also in other areas.

The second step is to turn the new physics theory into concrete predictions.
These should be sufficiently accurate, since the EFT operator effects
are often a small correction to the pure SM predictions. The predictions are provided by Monte Carlo generator codes, which produce event samples that are representative of the true particles momenta distributions. Accurate simulation of the detector response are further applied in order to obtain a representation of the distribution as a function of the variables that are actually observed in the experiment. It should be mentioned that this program occasionally fails. Namely it could be impossible for the Monte Carlo codes to provide a sufficiently accurate representation of all the components of the data distribution, for instance of reducible backgrounds from misidentification. In this case, the artificial Monte Carlo event sample should be supplemented with natural data collected in some control region, which model the missing component. We will not discuss this case explicitly, however it should be emphasized that our methodology would apply straightforwardly. Namely, the ``Monte Carlo'' samples we refer to in what follows might well not be the output of a Monte Carlo code, but rather have (partially) natural origin.

The final step, i.e.~the actual comparison of the predictions with the data, is often further split in two, by identifying suitable high-level observables (e.g.~cross sections in bins) that are particularly sensitive to the EFT operators. These observables can be measured in an experimental analysis that does not target the EFT explicitly, and compared with the EFT predictions at a later stage. This is convenient from the experimental viewpoint because the results are
model independent and thus potentially useful also to probe other new physics theories. If the measurements are performed at the truth (unfolded) level, this is also convenient for theorists because detector effects need not to be included in the predictions. However a strategy based on intermediate high-level observables is unavoidably suboptimal. It would approach optimality only if the fully differential distribution was measured for all the relevant variables, with sufficiently narrow binning. However there are often too many discriminating variables to measure their distribution fully differentially, and, even if this was feasible, one would not be able to predict accurately the cross section in too many bins. In this situation, the sensitivity to the presence (or absence) of the EFT operators could be strongly reduced and it could be impossible to disentangle the effect of different operators and resolve flat directions in the parameter space of the EFT Wilson coefficients.
One should thus switch to the direct comparison of the EFT with the data, by employing more sophisticated unbinned multivariate data analysis techniques.

Several multivariate methods have been developed and applied to the EFT or to similar problems, including Optimal Observables~\cite{Atwood:1991ka,Diehl:1993br}, the so-called ``Method of Moments''~\cite{Dunietz:1990cj,Dighe:1998vk,Banerjee:2019twi} and similar approaches (see e.g.~Ref.~\cite{Anderson:2013afp}) based on parametrizations of the scattering amplitude. The virtue of these techniques is that they are still based on high-level observables, making data/theory comparison simpler. The disadvantage is that they are intrinsically suboptimal and not systematically improvable towards optimality. 

A potentially optimal approach, which is closely analog to the one based on Machine Learning we employ in this paper, is the ``Matrix Element Method''~\cite{Kondo:1988yd,Artoisenet:2010cn,Fiedler:2010sg,Martini:2015fsa,Martini:2017ydu}. The main idea behind this construction is that optimal data analysis performances are unmistakably obtained by employing the likelihood ${\mathcal{L}}(c\,|{\cal{D}})$, i.e.~the probability density of the observed dataset ``${\cal{D}}$'' seen as a function of the free parameters ``$c$'' of the probability model. In our case, the free parameters coincide with the EFT
Wilson coefficients.\footnote{The notion of ``optimality'' can be made fully rigorous and quantitative, both when the purpose of the analysis is to measure the free parameters of the EFT and when it is to test the EFT hypothesis ($c\neq0$) against the SM ($c=0$) one, and both from a Bayesian and from a frequentist viewpoint. The case of a frequentist hypothesis test is discussed in more details below.} The LHC data consist of independent repeated measurements of the variable ``$x$'' that describes the kinematical configuration of each observed event. Therefore the likelihood factorizes and evaluating it requires only the knowledge of the distribution of $x$. More precisely, since we are interested in ${\mathcal{L}}(c\,|{\cal{D}})$ only up to an overall $c$-independent normalization, it is sufficient to know the ratio $r(x,c)$ between the density as a function of $c$ (and $x$) and the density at some fixed value $c={\overline{c}}$. The SM point, ${\overline{c}}=0$, can be conventionally chosen. 

It should be emphasized that extracting $r(x,c)$ is a highly non-trivial task, as nicely explained in Ref.~\cite{Brehmer:2018kdj} in terms of latent variables. The Monte Carlo generator code does of course implement an analytic point-by-point representation of the density (and, in turn, of $r$), which is however expressed in terms of abstract variables ``$z$''
and not of the variables $x$ that are actually observed. The analytic representation of $r$ in the $z$ variables can be used as a surrogate of $r(x,c)$ only if there is a faithful one-to-one correspondence between $z$ and $x$. This is typically the case at leading order, if showering and detector effects are small, and if there are not undetected particles. However it is sufficient to have neutrinos in the final state, or to include Next to Leading Order (NLO) corrections to spoil the
correspondence between $z$ and $x$. Showering, hadronization and detector effects also wash out the correspondence. In the Matrix Element Method, $r(x,c)$ is obtained by a phenomenological parameterization of these effects in terms of transfer functions that translate the knowledge of the density at the ``$z$'' level into the one at the ``$x$'' level. The free parameters of the phenomenological modeling of the transfer functions are fitted to Monte Carlo samples.

The Matrix Element Method is potentially optimal and improvable towards optimality. However it is not ``systematically'' improvable, in the sense that a more accurate reconstruction of $r(x,c)$ requires a case-by-case optimization of the transfer function modeling. With the alternative employed in this paper, based on the reconstruction of $r(x,c)$ using Machine Learning techniques rather than phenomenological modeling, systematic improvement is possible using bigger Neural Networks and larger training sets. Furthermore refining the reconstruction by including additional effects requires substantial effort and increases the computational complexity of the Matrix Element Method, while the complexity of the Machine Learning-based reconstruction is a priori independent of the degree of refinement of the simulations. Therefore it is important to investigate these novel techniques as an alternative and/or as a complement to the Matrix Element approach.

There is already a considerable literature on the reconstruction of $r(x,c)$ using Neural Networks~\cite{Cranmer:2015bka,Baldi:2016fzo,Stoye:2018ovl,Brehmer:2018hga,Brehmer:2018eca,Brehmer:2018kdj,Brehmer:2019xox} and several algorithms exist. Here we adopt the most basic strategy, mathematically founded on the standard Statistical Learning problem of classification (see Section~\ref{SC} for a brief review), which we improve by introducing the notion of ``Quadratic Classifier''. The relation between our methodology and the existing literature, the possibility of integrating it in other algorithms and to apply it to different problems is discussed in details in Section~\ref{PaC} and in the Conclusions. However it is worth anticipating that, unlike simulator-assisted techniques~\cite{Brehmer:2018hga}, the Quadratic Classifier only exploits Monte Carlo data samples (in the extended sense outlined above) and no other information on the data generation process. It can thus be used as it is with any Monte Carlo generator.

Apart from describing the Quadratic Classifier, the main aim of the paper is to investigate the potential impact of Machine Learning methods on LHC EFT searches, from two viewpoints. 

The first question we address is if and to what extent statistically optimal sensitivity to the presence or absence of the EFT operators can be achieved. In order to answer, a rigorous quantitative notion of optimality is defined by exploiting the Neyman--Pearson lemma~\cite{Neyman:1933wgr}, namely the fact that the ``best'' (maximum power at fixed size, in the standard terminology of e.g. Ref.~\cite{Tanabashi:2018oca}) frequentist test between two simple hypotheses is the one that employs the likelihood ratio as test statistic. By regarding the EFT at each given value of the $c$ Wilson coefficients as a simple hypothesis, to be compared with the SM $c=0$ hypothesis, we would thus obtain the strongest expected $95\%$ Confidence Level (CL) exclusion bounds on $c$ (when the SM is true) if the true distribution ratio $r(x,c)$ was available and used for the test. This can be compared with the bound obtained by employing the approximate ratio ${\widehat{r}}(x,c)$ reconstructed by the Neural Network, allowing us to quantify the approximation performances of the method in objective and useful terms.\footnote{It should be emphasized that we adopt this specific notion of ``optimality'' only because the frequentist hypothesis test between two simple hypotheses is relatively easy to implement in a fully rigorous manner. The reconstructed likelihood ratio could be employed for any other purpose and/or relying on asymptotic approximations using standard statistical techniques.} Of course, the exact $r(x,c)$ is not available in a realistic EFT problem, therefore the comparison can only be performed on a toy problem. In order to make it as close as possible to reality, our ``Toy'' problem is defined in terms of an analytical approximation of the differential cross section of the process of interest (i.e.~fully leptonic ZW, see below), implemented in a dedicated Monte Carlo generator.

The second aim of the paper is to quantify the potential gain in sensitivity of Machine Learning techniques, compared with the basic approach based on differential cross section measurements in bins. The associated production of a Z and a W boson decaying to leptons, at high transverse momentum ($p_{T,{\rm{Z}}}>300$~GeV) and with the total integrated luminosity of the High Luminosity LHC (HL-LHC), is considered for illustration. This final state has been selected to be relatively simple, but still described by a large enough number of variables to justify the usage of unbinned analysis techniques. Moreover it has been studied already quite extensively in the EFT literature~(see e.g.~Refs.~\cite{Falkowski:2015jaa,Green:2016trm,Butter:2016cvz,Franceschini:2017xkh,Panico:2017frx,Azatov:2017kzw,Azatov:2019xxn,Baglio:2019uty}) and a number of variables have been identified, including those associated with the vector bosons decay products~\cite{Duncan:1985ij,Hagiwara:1986vm,Panico:2017frx}, with the potential of improving the sensitivity to the EFT operators. 

The comparison with the binned analysis is performed on the Toy version of the problem mentioned above, on the exact tree-level (LO) modeling of the process and on NLO QCD plus parton showering Monte Carlo data. By progressively refining our modeling of the problem in these three stages, this comparison also illustrates the flexibility of the approach and the fact that increasingly sophisticated descriptions of the data are not harder for the machine to learn. This should be contrasted with the Matrix Element method, which would instead need to be substantially redesigned at each step. As an illustration, we will show that employing the analytical approximated distribution ratio, that was optimal on the Toy problem, leads to considerably worse performances than the Neural Network already at LO. At NLO the performances further deteriorate and the reach is essentially identical to the one of the binned analysis.

The rest of the paper is organized as follows. In Section~\ref{sec:teaching} we introduce the Quadratic Classifier as a natural improvement of the standard Neural Network classifier for cases, like the one of the EFT, where the dependence of the distribution ratio on the ``$c$'' parameters is known. The fully leptonic ZW process, the EFT operators we aim at probing and the relevant kinematical variables, are discussed in Section~\ref{sec:WZ}. The Toy, the LO and the NLO Monte Carlo generators employed in the analysis are also described. The first set of results, aimed at assessing the optimality of the Quadratic Classifier on the Toy data, are reported in Section~\ref{sec:IR}. The results obtained with the LO Monte Carlo are also discussed, showing the stability of the Neural Network performances as opposite to the degradation of the sensitivity observed with the Matrix Element and with the binned analysis methods. NLO results are shown in Section~\ref{sec:NLO}. We will see that the Quadratic Classifier methodology applies straightforwardly at NLO in spite of the fact that negative weights are present in the NLO Monte Carlo samples. The only complication associated with negative weights, which we discuss in Section~\ref{sec:tsdNLO}, is not related with the reconstruction of the ${\widehat{r}}(x,c)$ function by the Neural Network, but with the calculation of the distribution of the variable ${\widehat{r}}(x,c)$ itself, which is needed for the hypothesis test. All the technical details on the Neural Network design and training are summarized in Section~\ref{sec:IAV}, and our conclusions are reported in Section~\ref{sec:conc}. Appendices~\ref{app:A} and~\ref{app:B} contain the generalization of the Quadratic Classifier for an arbitrary number of Wilson coefficients and the proof of its asymptotic optimality.

\vspace{-5pt}
\section{Teaching new physics to a machine}\label{sec:teaching}

Consider two hypotheses, $H_0$ and $H_1$, on the physical theory that describes the distribution of the variable $x$. In the concrete applications of the following sections, $H_0$ will be identified with the SM EFT and $H_1$ with the SM theory. The statistical variable $x\in {\rm{X}}$ describes the kinematical configuration in the search region of interest X. In the following, $x$ will describe the momenta of $3$ leptons and the missing transverse momentum, subject to selection cuts. Each of the two hypotheses (after choosing, if needed, their free parameters) characterizes the distribution of $x$ completely. Namely they contain all the information that is needed to compute, in line of principle, the differential cross sections $d\sigma_0(x)$ and $d\sigma_1(x)$. The differential cross sections describe both the probability density function of $x$, which is obtained by normalization 
\beq
{\rm{pdf}}(x|H_{0,1})=\frac1{\sigma_{0,1}({\rm{X}})}\frac{d\sigma_{0,1}}{dx}\,,
\eeq
and the total number of instances of $x$ (i.e.~of events) that is expected to be found in the dataset, denoted as ${\rm{N}}({\rm{X}}|H_{0,1})$. This is equal to the cross section integrated on X and multiplied by the luminosity of the experiment, namely ${\rm{N}}({\rm{X}}|H_{0,1})={\rm{L}}\cdot\sigma_{0,1}({\rm{X}})$.

The total number of observed events follows a Poisson distribution. Hence for a given observed dataset ${\mathcal{D}}=\{x_i\}$, with ${\mathcal{N}}$ observed events, the $H_1/H_0$ log likelihood ratio reads
\beq\label{llr}
\lambda({\mathcal{D}})\equiv\log\frac{\mathcal{L}(H_1|{\mathcal{D}})}{\mathcal{L}(H_0|{\mathcal{D}})}={\rm{N}}({\rm{X}}|H_{0})-{\rm{N}}({\rm{X}}|H_{1})
- \sum\limits_{i=1}^{\mathcal{N}}\log\frac{d\sigma_0(x_i)}{d\sigma_1(x_i)}\,.
\eeq
The statistic $\lambda({\mathcal{D}})$ is known as the ``extended'' log likelihood ratio \cite{Cowan:1998ji}, and it is the central object for hypothesis testing ($H_0$ versus $H_1$) or for measurements (if $H_0$ contains free parameters), both from a Frequentist and from a Bayesian viewpoint. The ``N'' terms in eq.~(\ref{llr}) can be computed as Monte Carlo integrals. What is missing in order to evaluate $\lambda$ is thus the cross section ratio
\beq
r(x)\equiv\frac{d\sigma_0(x)}{d\sigma_1(x)}\,.
\eeq
This should be known locally in the phase space as a function of $x$.

The physical knowledge of the $H_0$ and $H_1$ models gets translated into Monte Carlo generator codes, which allow us to estimate $\sigma_{0,1}({\rm{X}})$ and to produce samples, ${\rm{S}}_{0,1}$, of artificial events following the ${\rm{pdf}}(x|H_{0,1})$ distributions. More precisely, the Monte Carlo generates weighted events \mbox{${\rm{e}}=(x_{\rm{e}},w_{\rm{e}})$}, with $x_{\rm{e}}$ one instance of $x$ and $w_{\rm{e}}$ the associated weight. If the $w_{\rm{e}}$'s are not all equal, $x_{\rm{e}}$ does not follow the pdf of $x$ and the expectation value of the observables $O(x)$ has to be computed as a weighted average. We choose the normalization of the weights such that they sum up to $\sigma_{0,1}({\rm{X}})$ over the entire sample
\beq
\sum\limits_{{\rm{e}}\in{\rm{S}}_{0,1}} w_{\rm{e}} = \sigma_{0,1}({\rm{X}})\,.
\eeq
With this convention, the weighted sum of $O(x_{\rm{e}})$ approximates the integral of \mbox{$O(x)\hspace{-1pt}\cdot\hspace{-1pt} d\sigma_{0,1}\hspace{-1.5pt}(x)$} on $x\in$ X. Namely
\beq\label{LS}
\sum\limits_{{\rm{e}}\in{\rm{S}}_{0,1}} w_{\rm{e}} O(x_{\rm{e}})\;\overset{\rm\sc{LS}}{\longrightarrow}\;\;\int_{x\in {\rm{X}}} \hspace{-4pt}d\sigma_{0,1}(x)\,O(x)=\sigma_{0,1}({\rm{X}})\,{\rm{E}}\left[O|H_{0,1}\right]\,,
\eeq
in the Large Sample ({\sc{LS}}) limit where ${\rm{S}}_{0,1}$ are infinitely large. We will see below how to construct an estimator ${\widehat{r}}(x)$ for $r(x)$ (or, in short, to fit $r(x)$) using finite ${\rm{S}}_{0}$ and ${\rm{S}}_{1}$ samples.

For tree-level Monte Carlo generators the previous formulas could be made simpler by employing unweighted samples where all the weights are equal. However radiative corrections need to be included for sufficiently accurate predictions, at least up to NLO in the QCD loop expansion. NLO generators can only produce weighted events, and some of the events have a negative weight. Therefore the NLO Monte Carlo samples cannot be rigorously interpreted as a sampling of the underlying distribution. However provided they consistently obey the {\sc{LS}} limiting condition in eq.~(\ref{LS}), they are equivalent to ordinary samples with positive weights for most applications, including the one described below.

\subsection{The Standard Classifier}\label{SC}

The estimator ${\widehat{r}}(x)$ can be obtained by solving the most basic Machine Learning problem, namely supervised classification with real-output Neural Networks (see Ref.~\cite{NNL} for an in-depth mathematical discussion). One considers a Neural Network acting on the kinematical variables and returning $f(x)\in(0,1)$. This is trained on the two Monte Carlo samples by minimizing the loss function 
\beq\label{LF}
L[f(\cdot)]=\sum\limits_{{{\rm{e}}\in{\rm{S}}_{0}}}w_{\rm{e}}[f(x_{\rm{e}})]^2 + \sum\limits_{{{\rm{e}}\in{\rm{S}}_{1}}} w_{\rm{e}}[1-f(x_{\rm{e}})]^2\,,
\eeq
with respect to the free parameters (called weights and biases) of the Neural Network. The trained Neural Network, ${\widehat{f}}(x)$, is in one-to-one correspondence with ${\widehat{r}}(x)$, namely
\beq
{\widehat{f}}(x) = \frac1{1+ {\widehat{r}}(x)}\;\;\;\Leftrightarrow\;\;\;  {\widehat{r}}(x)=\frac1{{\widehat{f}}(x)}-1\,.
\eeq
The reason why ${\widehat{r}}(x)$, as defined above, approximates $r(x)$ is easily understood as follows. If the Monte Carlo training data are sufficiently abundant, the loss function in eq.~(\ref{LF}) approaches its Large Sample limit and becomes
\beq\label{LFL}
L[f(\cdot)]\;\overset{\rm\sc{LS}}{\longrightarrow}\;\;\int_{x\in{{\rm{X}}}} \hspace{-4pt}d\sigma_{0}(x) [f(x)]^2+\int_{x\in{{\rm{X}}}}  \hspace{-4pt} d\sigma_{1}(x) [1-f(x)]^2\,.
\eeq
Furthermore if the Neural Network is sufficiently complex (i.e.~contains a large number of adjustable parameters) to be effectively equivalent to an arbitrary function of $x$, the minimum of the loss can be obtained by variational calculus. By setting to zero the functional derivative of $L$ with respect to $f$ one immediately finds
\beq
{\widehat{f}}(x) \simeq\frac{d\sigma_1(x)}{d\sigma_1(x)+d\sigma_0(x)}= \frac1{1+ {{r}}(x)}\;\;\;\Rightarrow\;\;\;{\widehat{r}}(x)\simeq r(x)\,.
\eeq
The same result holds for other loss functions such as the standard Cross-Entropy, which has been found in Ref.~\cite{Stoye:2018ovl} to have better performances for EFT applications, or the more exotic ``Maximum Likelihood'' loss~\cite{DAgnolo:2018cun}, which is conceptually appealing because of its connection with the Maximum Likelihood principle. We observed no strikingly different performances with the various options, but we did not investigate this point in full detail. In what follows we stick to the quadratic loss in eq.~(\ref{LF}).

The simple argument above already illustrates the two main competing aspects that control the performances of the method and its ability to produce a satisfactory approximation of $r(x)$. One is that the Neural Network should be complex in order to attain a configuration that is close enough to the (absolute) minimum, $f(x)=1/(1+r(x))$, of the loss functional in eq.~(\ref{LFL}). In ordinary fitting, this is nothing but the request that the fit function should contain enough adjustable parameters to model the target function accurately. On the other hand if the Network is too complex, it can develop sharp features, while we are entitled to take the Large Sample limit in eq.~(\ref{LFL}) only if $f$ is a smooth enough function of $x$. Namely we need $f$ to vary appreciably only in regions of the X space that contain enough Monte Carlo points. Otherwise the minimization of eq.~(\ref{LF}) brings $f$ to approach zero at the individual points that belong to ${\rm{S}}_{0}$ sample, and to approach one at those of the ${\rm{S}}_{1}$ sample. This phenomenon, called overfitting, makes that for a given finite size of the training sample, optimal performances are obtained by balancing the intrinsic approximation error of the Neural Network against the complexity penalty due to overfitting. A third aspect, which is extremely important but more difficult to control theoretically, is the concrete ability of the training algorithm to actually reach the global minimum of the loss function in finite time. This requires a judicious choice of the minimization algorithm and of the Neural Network activation functions. 

The problem of fitting ${{r}}(x)$ is mathematically equivalent to a classification problem. A major practical difference however emerges when considering the level of accuracy that is required on ${\widehat{r}}(x)$ as an approximation of ${{r}}(x)$. Not much accuracy is needed for ordinary classification, because ${\widehat{r}}(x)$ (or, equivalently, ${\widehat{f}}(x)$) is used as a discriminant variable to distinguish instances of $H_0$ from instances of $H_1$ on an event-by-event basis. Namely, one does not employ ${\widehat{r}}(x)$ directly in the analysis of the data, but a thresholded version of ${\widehat{r}}(x)$ that isolates regions of the X space that are mostly $H_0$-like ($r$ is large) or $H_1$-like ($r$ is small). Some correlation between ${\widehat{r}}(x)$ and $r(x)$, such that ${\widehat{r}}(x)$ is large/small when ${{r}}(x)$ is large/small, is thus sufficient for good classification performances. Furthermore the region where $r(x)\simeq1$ is irrelevant for classification.

The situation is radically different in our case because the EFT operators are small corrections to the SM. The regions where the EFT/SM distribution ratio is close to one cover most of the phase-space, but these regions can contribute significantly to the sensitivity if they are highly populated in the data sample. Mild departures of $r(x)$ from unity should thus be captured by ${\widehat{r}}(x)$, with good accuracy relative to the magnitude of these departures. Obviously the problem is increasingly severe when the free parameters of the EFT (i.e.~the Wilson coefficients ``$c$'') approach the SM value $c=0$ and $r(x)$ approaches one. On the other hand it is precisely when $c$ is small, and the EFT is difficult to see, that a faithful reconstruction of $r(x)$ would be needed in order to improve the sensitivity of the analysis.

\subsection{The Quadratic Classifier}\label{PaC}

Barring special circumstances, the EFT prediction for the differential cross section is a quadratic polynomial in the Wilson coefficients.\footnote{The only exception is when the relevant EFT effects are modifications of the SM particles total decay widths. Also notice that the cross section is quadratic only at the leading order in the EFT perturbative expansion, which is however normally very well justified since the EFT effects are small. Higher orders could nevertheless be straightforwardly included as higher order polynomial terms.} If a single operator is considered, so that a single free parameter $c$ is present and the SM corresponds to the value $c=0$, the EFT differential cross section reads
\beq\label{QP}
d\sigma_0(x;c)=d\sigma_1(x)\left\{[1+c\,\alpha(x)]^2+[c\,\beta(x)]^2\right\}\,,
\eeq
where $\alpha(x)$ and $\beta(x)$ are real functions of $x$. An estimator ${\widehat{r}}(x,c)$ for the distribution ratio in the entire Wilson coefficients parameters space could thus be obtained as 
\beq
{\widehat{r}}(x,c)=[1+c\,\widehat\alpha(x)]^2+[c\,\widehat\beta(x)]^2\,,
\eeq
from estimators $\widehat\alpha(x)$ and $\widehat\beta(x)$ of the coefficient functions $\alpha(x)$ and $\beta(x)$. Notice that eq.~(\ref{QP}) parametrizes, for generic $\alpha(x)$ and $\beta(x)$, the most general function of $x$ and $c$ which is quadratic in $c$, which is always positive (like a cross section must be) and which reduces to the SM cross section for $c=0$. The equation admits a straightforward generalization for an arbitrary number of $c$ parameters, which we work out in Appendix~\ref{app:A}.

The estimators $\widehat\alpha(x)$ and $\widehat\beta(x)$ are obtained as follows. We first define a function $f(x;c)\in(0,1)$, in terms of two neural networks ${\rm{n}}_\alpha$ and ${\rm{n}}_\beta$ with unbounded output (i.e.~${\rm{n}}_{\alpha,\beta}\in(-\infty,+\infty)$ up to weight-clipping regularization), with the following dependence on $c$:
\beq\label{PC}
f(x,c)\equiv\frac1{1+[1+c\,{\rm{n}}_{\alpha}(x)]^2+[c\,{\rm{n}}_{\beta}(x)]^2}\,.
\eeq
Next, we consider a set ${\mathcal{C}}=\{c_i\}$ of values of $c$ and we generate the corresponding EFT Monte Carlo samples ${\rm{S}}_{0}(c_i)$. At least two distinct values of $c_i\neq0$ need to be employed, however using more than two values is beneficial for the performances. Monte Carlo samples are also generated for the $H_1$ (i.e.~$c=0$) hypothesis, one for each of the ${\rm{S}}_{0}(c_i)$ samples. These are denoted as ${\rm{S}}_{1}(c_i)$ in spite of the fact that they are all generated according to the same $c=0$ hypothesis. The samples are used to train the ${\rm{n}}_{\alpha,\beta}$ Networks, with the loss function
\beq\label{LFPC}
L[{\rm{n}}_{\alpha}(\cdot),{\rm{n}}_{\beta}(\cdot)]=\sum\limits_{c_i\in{\mathcal{C}}}\left\{
\sum\limits_{{{\rm{e}}\in{\rm{S}}_{0}}(c_i)}w_{\rm{e}}[f(x_{\rm{e}},c_i)]^2 + \sum\limits_{{{\rm{e}}\in{\rm{S}}_{1}}(c_i)} w_{\rm{e}}[1-f(x_{\rm{e}},c_i)]^2
\right\}
\,.
\eeq
We stress that in the second term in the curly brackets, the function $f$ is evaluated on the same value of $c=c_i$ that is employed for the generation of the ${\rm{S}}_{0}(c_i)$ Monte Carlo sample which we sum over in the first term.

By taking the Large Sample limit for the loss function as in eq.~(\ref{LFL}), differentiating it with respect to ${\rm{n}}_{\alpha}$ and ${\rm{n}}_{\beta}$ and using the quadratic condition~(\ref{QP}), it is easy to show that the trained Networks $\widehat{\rm{n}}_{\alpha}$ and $\widehat{\rm{n}}_{\beta}$ approach $\alpha$ and $\beta$, respectively. Namely
\beq
\widehat\alpha(x)\equiv\widehat{\rm{n}}_{\alpha}(x)\simeq\alpha(x)\,,\;\;\;\;\; \widehat\beta(x)\equiv\widehat{\rm{n}}_{\beta}(x)\simeq\beta(x)\,.
\eeq
More precisely, by taking the functional derivative one shows that the configuration ${\rm{n}}_{\alpha}=\alpha$ and ${\rm{n}}_{\beta}=\beta$ is a local minimum of the loss in the Large Sample limit. It is shown in Appendix~\ref{app:B} that this is actually the unique global minimum of the loss.

It is simple to illustrate the potential advantages of the Quadratic Classifier, based on the analogy with the basic binned histogram approach to EFT searches. In that approach, the X space is divided in bins and the likelihood ratio is approximated as a product of Poisson distributions for the countings observed in each bin. Rather than ${\widehat{r}}(x,c)$, the theoretical input required to evaluate the likelihood are estimates ${\widehat\sigma}_0({\rm{b}};c)$ for the cross sections integrated in each bin ``${\rm{b}}$''. Employing the Standard Classifier approach to determine ${\widehat{r}}(x,c)$ would correspond in this analogy to compute ${\widehat\sigma}_0({\rm{b}};c)$ for each fixed value of $c$ using a dedicated Monte Carlo simulation. By scanning over $c$ on a grid, ${\widehat\sigma}_0({\rm{b}};c)$ would be obtained by interpolation. Every EFT practitioner knows that this is a is very demanding and often unfeasible way to proceed. Even leaving aside the computational burden associated with the scan over $c$, the problem is that the small values of $c$ (say, $c=\overline{c}$) we are interested in probing typically predict cross sections that are very close to the SM value and it is precisely the small relative difference between the EFT and the SM predictions what drives the sensitivity. A very small Monte Carlo error, which in turn requires very accurate and demanding simulations, would be needed in order to be sensitive to these small effects. In the Standard Classifier method, the counterpart of this issue is the need of generating very large samples for training the Neural Network. Furthermore, this should be done with several values of $c$ for the interpolation. This approach is computationally demanding even when a single Wilson coefficient is considered, and it becomes rapidly unfeasible if $c$ is a higher-dimensional vector of Wilson coefficients to be scanned over.

The strategy that is normally adopted in standard binned analyses is closely analog to a Quadratic Classifier. One enforces the quadratic dependence of ${\sigma}_0({\rm{b}};c)$ on $c$ as in eq.~(\ref{QP}), and estimates the three polynomial coefficients (i.e.~the SM cross section and the analog of $\alpha$ and $\beta$) in each bin by a $\chi^2$ fit to ${\widehat\sigma}_0({\rm{b}};c)$, as estimated from the Monte Carlo simulations for several values of $c$. The values of $c$ used for the fit are much larger that the reach of the experiment $c=\overline{c}$, so that their effects are not too small and can be captured by the Monte Carlo simulation. The Quadratic Classifier works in the exact same way. It can learn $\widehat\alpha(x)$ and $\widehat\beta(x)$ using training samples generated with large values of $c$, for which the difference between the ${\rm{S}}_{0}(c)$ and ${\rm{S}}_{1}(c)$ is sizable. The training can thus recognize this difference, producing accurate estimates of $\widehat\alpha(x)$ and $\widehat\beta(x)$. This accurate knowledge results in an accurate estimate of ${\widehat{r}}(x,c)$ and of its departures from unity even at the small value $c=\overline{c}$, because our method exploits the exact quadratic relation in eq.~(\ref{QP}). 

It should be noted that the ``Quadratic Classifier'' introduced in eq.~(\ref{PC}) is ``Parametrized'' in the sense that it encapsulates the dependence on the $c$ parameters, but it is the exact opposite of the Parametrized Neural Network (or Parametrized Classifier) of Ref.~\cite{Baldi:2016fzo}. In that case, the Wilson Coefficient $c$ is given as an input to the Neural Network, which acts on an enlarged $(x,c)$ features space. The purpose is to let the Neural Network learn also the dependence on $c$ of the distribution ratio in cases where this is unknown. Here instead we want to enforce the quadratic dependence of the distribution ratio on $c$, in order to simplify the learning task. 

An alternative strategy to exploit the analytic dependence on $c$ is the one based on ``morphing''~\cite{Brehmer:2018eca}. Morphing consists in selecting one point in the parameter space for each of the coefficient functions that parametrize $d\sigma_0(x;c)$ as a function of $c$, and expressing $d\sigma_0(x;c)$ as a linear combination of the cross-sections computed at these points. For instance, a total of $3$ ``morphing basis points'', $c_{1,2,3}$, are needed for a single Wilson coefficient and quadratic dependence, and $d\sigma_0(x;c)$ is expressed as a linear combination of $d\sigma_0(x;c_{1,2,3})$. This rewriting can be used to produce two distinct learning algorithms. 

The first option is to learn the density ratios $d\sigma_0(x;c_{1,2,3})/d\sigma_1(x)$ individually (one-by-one or simultaneously), by using training data generated at the morphing basis points $c_{1,2,3}$, and to obtain ${\widehat{r}}(x,c)$ using the morphing formula. In the analogy with ordinary binning, this would correspond to extracting the dependence on $c$ of the cross-sections by a quadratic interpolation of ${\widehat\sigma}_0({\rm{b}};c_{1,2,3})$ at the selected points. Of course it is possible to reconstruct the cross sections accurately also by using $3$ very accurate simulations, rather than fitting less accurate simulations at several points. However a judicious choice of the values of $c_{1,2,3}$ is essential for a proper reconstruction of the quadratic and of the linear term of the polynomial. For the former, it is sufficient to take $c$ very large, but for the latter a value of $c$ should be selected that is neither too large, such that the quadratic term dominates by too many orders of magnitude, nor too small such that the constant SM term dominates. Notice that the optimal $c$ depends on the analysis bin because the EFT effects relative to the SM (and the relative importance of the quadratic and linear terms) can be vastly different in different regions of the phase space. With ``plain'' morphing as described above, we are obliged to employ only few values of $c$, which might not be enough to cover the entire phase space accurately. With the Quadratic Classifier instead, all values of $c$ that are useful to learn the distribution in some region of the phase space (see e.g. eq.~(\ref{tranv})) can be included simultaneously in the training set. 

Alternatively, one can use the morphing formula in place of eq.~(\ref{QP}), producing a different parameterization of the classifier than the one in eq.~(\ref{PC}), to be trained with values of the parameters that are unrelated with the morphing basis points. The parametrization employed in the Quadratic Classifier is arguably more convenient, as it is simpler, universal and bounded to $f\in (0,1)$ interval owing to the positivity of eq.~(\ref{QP}). Importantly, also the condition ${\widehat{r}}(x,0)\equiv1$ is built-in in the Quadratic Classifier. However this could be enforced in the morphing parameterization as well by selecting $c=0$ as one of the basis points. If this is done, we do not expect~\footnote{Provided that the possibility of having $f$ outside the $(0,1)$ interval is not a problem when training, for instance, with the cross-entropy loss function.} a degradation of the performances if employing the morphing-based parametrization rather then ours. Indeed, we believe that the key of the success of the Quadratic Classifier that we observe in this paper stems from the appropriate choice of the values of $c$ used for training, and not from the specific parametrization we employ. The non-optimal performances of the morphing strategy observed in Ref.~\cite{Brehmer:2018eca} (on a different process than the one we study) are probably to be attributed to a non-optimal choice. Further investigations on this aspect are beyond the scope of the present paper.

\section{Fully leptonic ZW}\label{sec:WZ}

Consider ZW production at the LHC with leptonic decays, namely $Z \rightarrow \ell^+ \ell^-$ and $W \rightarrow \ell \nu$, where $\ell = e, \mu$. As explained in the Introduction, this is arguably the simplest process, of established EFT relevance, where a multivariate approach is justified and potentially improves the sensitivity. We focus on the high-energy tail of the process, with a selection cut on the transverse momentum of the Z-boson, $p_{T,{\rm{Z}}}>300$~GeV, because of two independent reasons. First, because at high energy we can approximate the differential cross section analytically and define a realistic enough Toy problem to assess the optimality of the method. Second, because at high-energy the statistics is sufficiently limited (less than $5\times 10^3$ expected events at the HL-LHC, including both W charges) to expect  systematic uncertainties not to play a dominant role. The reach we will estimate in the $p_{T,{\rm{Z}}}>300$~GeV region, on purely statistical bases, should thus be nearly realistic.

The high-energy regime, in spite of the relatively limited statistics, is the most relevant one to probe those EFT operators that induce energy-growing corrections to the SM amplitudes. There are only two CP-preserving and flavor-universal operators in the ZW channel that induce quadratically energy-growing terms, namely \footnote{We use the definition $H^ \dagger {\scriptstyle \overleftrightarrow{\rule{0pt}{.75em}}}\hspace{-.95em}{D}_\mu H = H^\dagger D_\mu H - (D_\mu H)^\dagger H$.}
\beq
{\cal O}_{\varphi q}^{(3)} = G^{(3)}_{\varphi q} \left(\overline Q_L \sigma^a \gamma^\mu Q_L\right)
(i H^ \dagger {\scriptstyle \overleftrightarrow{\rule{0pt}{.75em}}}\hspace{-.95em}{D}_\mu H)\,,\qquad\quad
{\cal O}_{W} = G_{W} \varepsilon_{abc} {W^{a\,\nu}_{\mu}} {W^{b\,\rho}_{\nu}} {W^{c\,\mu}_\rho}\,.
\eeq
We thus focus on these operators in our analysis.

Both ${\cal O}_{\varphi q}^{(3)}$ and ${\cal O}_{W}$ contribute to the ZW production amplitudes with quadratically energy-growing terms of order $G\cdot{s}$, where $s$ is the center-of-mass energy squared of the diboson system. However the way in which this energy growth manifests itself in the cross section is rather different for the two operators (see e.g.~Refs.~\cite{Panico:2017frx,Franceschini:2017xkh}). The ${\cal O}_{\varphi q}^{(3)}$ operator mainly contributes to the ``$00$'' helicity amplitude, in which the gauge bosons are longitudinally polarized. The SM amplitude in  this channel is sizable and has a constant behavior with $s$ at high energy. As a consequence, a sizable quadratically-growing interference term between the SM and the BSM amplitudes is present in the cross section. This happens even at the ``inclusive'' level, i.e.~when only the hard scattering variables describing ZW production (and not the decay ones) are measured.

On the contrary, the ${\cal O}_{W}$ operator induces quadratically-growing contributions only in the transverse polarization channels with equal helicity for the two gauge bosons (namely, $++$ and $--$). In the SM this channel is very suppressed at high energy, since its amplitude decreases as $m_W^2/s$. Therefore in inclusive observables the interference between ${\cal O}_{W}$ and the SM does not grow with the energy and is very small. In order to access (or ``resurrect''~\cite{Panico:2017frx})  the interference, which is the dominant new physics contribution since the Wilson coefficient of the operator is small, the vector bosons decay variables must be measured. We thus expect that the sensitivity to ${\cal O}_{W}$ will benefit more from an unbinned multivariate analysis technique than the one on ${\cal O}_{\varphi q}^{(3)}$.

\begin{figure}
\centering
\includegraphics[width=.65\textwidth]{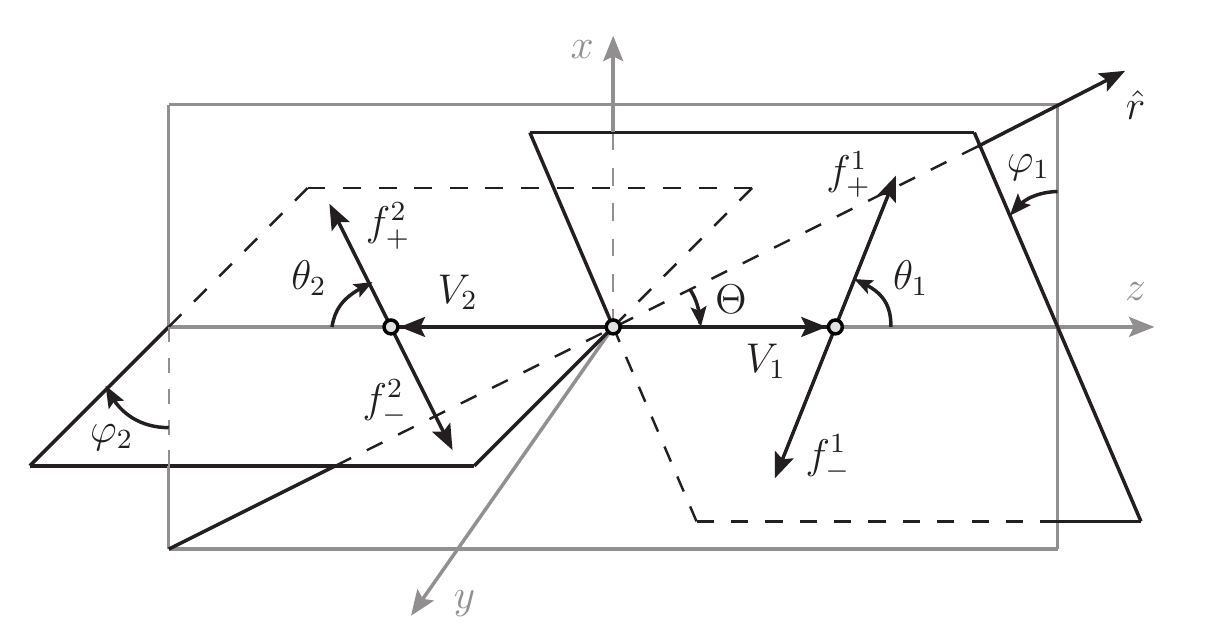}
\caption{The kinematical variables in the ``special' coordinate frame~\cite{Panico:2017frx}.}\label{fig:decay_angles}
\end{figure}

The relevant kinematical variables that characterize the four-leptons final state are defined as in Ref.~\cite{Panico:2017frx} and depicted in Figure~\ref{fig:decay_angles}, where $V_1$ is identified with the Z and $V_2$ with the W boson. The figure displays the kinematics in the rest frame of the ZW system, obtained from the lab frame by a boost along the direction of motion (denoted as ${\hat{r}}$ in the figure) of the ZW pair, followed by a suitable rotation that places the Z
along the positive $z$ axis and ${\hat{r}}$ on the $x$-$z$ plane with positive $x$ component. The ``inclusive'' variables associated with ZW production are the center-of-mass energy squared $s$ and $\Theta\in[0,\pi]$, which is defined as the angle between ${\hat{r}}$  and the Z-boson momentum. The decay kinematics is described by the polar and azimuthal decay angles $\theta_{1,2}$ and $\varphi_{1,2}$. The latter angles are in the rest frame of each boson and they are defined as those of the final fermion or anti-fermion with helicity $+1/2$ (e.g.~the $\ell^+$ in the case of a $W^+$ and the $\overline{\nu}$ for a $W^-$), denoted as $f_+^{1,2}$ in the figure.\footnote{The correct definition of $\varphi_2$ appears in version four of Ref.~\cite{Panico:2017frx}.} The remaining variables that are needed to characterize the four leptons completely are weakly sensitive to the presence of the EFT operators and can be ignored, with the exception of the total transverse momentum of the ZW system, $p_{T,{\rm{ZW}}}$, which is a useful discriminant at NLO~\cite{Franceschini:2017xkh}. 

The variables described above are useful for the theoretical calculation of the cross section, but they cannot be used for our analysis because they are not experimentally accessible. The ``measured'' variables we employ are defined as follows. First, since we do not measure the neutrino (longitudinal) momentum, this needs to be reconstructed by imposing the on-shell condition for the W. The reconstructed neutrino momentum, rather than the true one, is used to define the kinematical variables and in particular $s$ and $\Theta$. Moreover, since we do not measure the helicity of the fermions but only their charge, the decay angles of the Z boson, denoted as $\theta_Z$ and $\varphi_Z$, are defined in terms of the charge-plus lepton rather than of the helicity plus lepton. Depending on the (unobserved) leptons helicities these angles are either equal to $\theta_{1}$ and $\varphi_{1}$, or to $\pi-\theta_{1}$ and $\varphi_{1}+\pi$, respectively. The W decay angles, defined in terms of the lepton or the reconstructed neutrino depending on the charge of the W as previously explained, are denoted as  $\theta_W$ and $\varphi_W$. In summary, the variables we employ in the analysis are
\beq\label{feat}
\{ s,\,\Theta,\, \theta_W,\, \varphi_W,\, \theta_Z,\, \varphi_Z,\, p_{T,{\rm{ZW}}}  \}\,,
\eeq
where of course $p_{T,{\rm{ZW}}}$ is non-vanishing only at NLO.

The on-shell condition for the W boson has no real solution if the W-boson transverse mass is larger that the W pole mass $m_W$. The neutrino is reconstructed in this case by assuming that the neutrino rapidity is equal to the one of the lepton. If instead the transverse mass is smaller than $m_W$, the condition has two distinct real solutions, each of which produces a different reconstructed kinematics. For our analysis we picked one of the two solutions at random on an event-by-event basis, while for the analysis of the actual data it would be arguably more convenient to duplicate the kinematical variables vector by including both solutions. Nothing changes in the discussion that follows 
if this second option is adopted. 

\subsection{Analytic approximation}\label{sec:anap}

At the tree-level order, and based on the narrow-width approximation for the decays, it is easy to approximate the cross section analytically in the high-energy regime. The crucial simplification is that the reconstructed $3$-momentum of the W boson (with any of the two solutions for the neutrino) becomes exact when the W is boosted, so that the reconstructed $\Theta$ and $s$ variables approach the ``true'' ones of Figure~\ref{fig:decay_angles}. Notice that $\Theta$ is the angle between the Z and the direction of motion of the ZW system in the lab frame, which corresponds at tree-level to the direction of motion of the most energetic incoming parton. In the kinematical region we are interested in, the (valence) quark is more energetic than the anti-quark in more than $80\%$ of the events. Therefore we can identify $\Theta$ as the angle between the Z and the $u$ quark or the $d$ quark in the $u{\overline{d}}\to ZW^+$ and $d{\overline{u}}\to ZW^-$ processes, respectively. 

With these identifications, the non-vanishing on-shell helicity amplitudes ${\cal M}_{h_Zh_W}$ for the hard scattering process $u{\overline{d}}\to ZW^+$, at leading order in the high-energy expansion, read
\begin{eqnarray}\label{eq:amplitudes}
&\displaystyle{\cal M}_{00} = - \frac{g^2\sin \Theta}{2 \sqrt{2}} - \sqrt{2} G_{\varphi q}^{(3)}  s \sin \Theta\,,\qquad\quad
{\cal M}_{++}={\cal M}_{--}= \frac{3 g c_{\rm{w}} G_W  s \sin \Theta}{\sqrt{2}}\,,&\\
&\displaystyle{\cal M}_{-+} = - \frac{g^2(s_{\rm{w}}^2- 3 \, c_{\rm{w}}^2 \cos \Theta)}{3 \sqrt{2} c_{\rm{w}}} \cot \frac{\Theta}{2}\,,\qquad\quad
{\cal M}_{+-} = \frac{g^2(s_{\rm{w}}^2 - 3 c_{\rm{w}}^2 \cos \Theta)}{3 \sqrt{2} c_{\rm{w}}} \tan \frac{\Theta}{2}\,,&\nonumber
\end{eqnarray}
where $g$ is the SU$(2)_L$ coupling, $c_{\rm{w}}$ and $s_{\rm{w}}$ are the cosine and the sine of the Weak angle. An overall factor equal to the cosine of the Cabibbo angle has not been reported for shortness. The amplitudes for the $d{\overline{u}}\to ZW^-$ process can be obtained from the ones above with the formal substitutions $\Theta \rightarrow - \Theta$ and $s_{\rm{w}}^2 \rightarrow - s_{\rm{w}}^2$. The amplitudes are non-vanishing only for left-chirality initial quarks. Notice that the above formulas depend on the conventions in the definition of the wave-function of the external particles, and that these conventions must match the ones employed in the decay amplitude for the consistency of the final results. The wave-function reported in Ref.~\cite{Cuomo:2019siu} are employed.

Let us now turn to the vector bosons decays. The decay amplitudes assume a very simple form in terms of the $\theta=\theta_{1,2}$ and $\varphi=\varphi_{1,2}$ variables, namely
\begin{equation}
{\cal A}_{h} = - \sqrt{2} g_{V} m_{V} e^{i h \varphi} d_{h}(\theta)\,,
\end{equation}
where $h$ is the helicity of the decaying vector boson ($V=V_{1,2}=Z,W$) and $d_{h}(\theta)$ are the Wigner $d$-functions. The overall coupling factor $g_{V}$ depends on the nature of the boson and, in the case of the Z, on the electric charge of the helicity-plus fermion it decays to. Specifically, $g_{W}=g/\sqrt{2}$ for the W, $g_{Z}=g_L=-g(1-2\,s_{\rm{w}}^2)/2c_{\rm{w}}$ if the Z decays to an helicity-plus $\ell^+$ and $g_{Z}=g_R=g\,s_{\rm{w}}^2/c_{\rm{w}}$ if the Z decays to an helicity-plus $\ell^-$. The two options for the helicity (which are physically distinct) correspond to two terms in the cross section. In the first one the Z decay amplitude is evaluated with the $g_L$ coupling, with $\theta=\theta_1=\theta_Z$ and $\varphi=\varphi_1=\varphi_Z$. In the second one we employ $g_R$, $\theta=\theta_1=\pi-\theta_Z$ and $\varphi=\varphi_1=\varphi_Z+\pi$. There is no helicity ambiguity in the W-boson decay angles. However the reconstruction of the azimuthal decay angle is exact in the high-energy limit only up to a twofold ambiguity~\cite{Panico:2017frx}. Namely the reconstructed $\varphi_W$ approaches $\varphi_1$ on one of the two solutions for the neutrino (and we do not know which one), and $\pi-\varphi_1$ on the other. Since we are selecting one solution at random, we should average the cross section over the two possibilities $\varphi=\varphi_2=\varphi_W$ and $\varphi=\varphi_2=\pi-\varphi_W$ for the W azimuthal angle. The polar angle is instead $\theta=\theta_2=\theta_W$ in both cases.

Production and decay are conveniently combined using the density matrix notation. We define the hard density matrix
\begin{equation}
d \rho^{\rm hard}_{h_Z^{\phantom\prime} h_W^{\phantom\prime} h'_Z h'_W} =\frac1{24\,s}{\cal M}_{h_Z^{\phantom\prime} h_W^{\phantom\prime}} ({\cal M}_{h_Z^{\prime} h_W^{\prime}})^*\, d \Phi_{\rm{ZW}} \,,
\end{equation}
where $d \Phi_{\rm{ZW}}$ is the two-body phase space and the factor $1/24s$ takes care of the flux and of the averages over the colors and the helicities of the initial quarks. The decay processes are instead encoded into decay density matrices. The one for the Z-boson, including the sum over the $\ell^\pm$ helicities as previously explained, reads
\begin{equation}
d \rho^{Z}_{h_Z^{\phantom\prime} h'_Z} = \frac{1}{2m_Z\Gamma_Z} \left[
\left.{\cal A}_{h_Z^{\phantom{\prime}}}^{\phantom{*}} {\cal A}^*_{h'_Z}\right|_{g_L,\theta_Z,\varphi_Z} +
\left.{\cal A}_{h_Z^{\phantom{\prime}}}^{\phantom{*}} {\cal A}^*_{h'_Z}\right|_{g_R,\pi-\theta_Z,\varphi_Z+\pi} 
 \right] d \Phi_{\ell^+ \ell^-} \,,
\end{equation}
where $\Gamma_Z$ is the Z decay width. For the W, since we average on the neutrino reconstruction ambiguity, we have 
\begin{equation}
d \rho^{W}_{h_W^{\phantom\prime} h'_W} = \frac{1}{2m_W\Gamma_W} \frac12 \left[
\left.{\cal A}_{h_W^{\phantom{\prime}}}^{\phantom{*}} {\cal A}^*_{h^\prime_W}\right|_{\frac{g}{\sqrt{2}},\theta_W,\varphi_W} +
\left.{\cal A}_{h_W^{\phantom{\prime}}}^{\phantom{*}} {\cal A}^*_{h'_W}\right|_{\frac{g}{\sqrt{2}},\theta_W,\pi-\varphi_W} 
 \right] d \Phi_{\ell\nu} \,.
\end{equation}

The complete partonic differential cross section is finally simply given by
\begin{equation}\label{acs}
d \widehat\sigma = 4 \sum d \rho^{\rm hard}_{h_Z^{\phantom\prime} h_W^{\phantom\prime} h'_W h'_Z} d \rho^{Z}_{h_Z^{\phantom\prime} h'_Z}
d \rho^{W}_{h_W^{\phantom\prime} h'_W}\,,
\end{equation}
where the sum is performed on the four helicity indices and the factor of $4$ takes into account the decay channels into electrons and muons.

\subsection{Monte Carlo Generators}\label{mcgen}

For our analysis we use three Monte Carlo generators, of increasing accuracy.

The first one is the Toy generator that implements the analytic approximation of the cross section in eq.~(\ref{acs}), with the hard amplitudes expanded up to order $G\cdot s$ in the EFT contribution and up to order $s^0$ in the SM term, as in eq.~(\ref{eq:amplitudes}). This implies, in particular, that in the Toy Monte Carlo all the mixed transverse/longitudinal helicity channels vanish exactly, that only the $\pm\mp$ and $00$ channels are retained in the SM and that new physics is just in the $00$ and $\pm\pm$ channels for ${\cal O}_{\varphi q}^{(3)}$ and ${\cal O}_{W}$, respectively. The Toy Monte Carlo employs
a simple fit to the ($u\overline{d}$ or $d\overline{u}$) parton luminosities obtained from the nCTEQ15~\cite{Kusina:2015vfa} PDF set (implemented through the {\tt{ManeParse}}~\cite{Clark:2016jgm} Mathematica package). The variable $s$ is sampled according to the parton luminosity, while all the other variables are sampled uniformly. The cut $p_{T,Z}=\sqrt{s}/2\sin\Theta>300$~GeV is implemented at generation level. Since the analytical distribution is extremely fast to evaluate, this basic approach is sufficient to obtain accurate Monte Carlo integrals and large unweighted event samples in a very short time.

The second generator is {\sc MadGraph}~\cite{Alwall:2014hca} at LO, with the EFT operators implemented in the UFO model of Ref.~\cite{SMEFTMG}. We simulate the $2\to4$ process  $p p \rightarrow \mu^+ \mu^- e\, \nu_e$, with the Z and the W decaying to opposite flavor leptons for a simpler reconstruction, and multiply the resulting cross section by $4$. The cut on $p_{T,{\rm{Z}}}$, defined as the sum of the $\mu^+$ and $\mu^-$ momenta, is imposed at generation level, as well as the cuts
\begin{equation}\label{eq:cuts_inv_mass}
m_{T,e\nu} \leq 90\,{\rm GeV}\,, \qquad \quad
70\,{\rm GeV} \leq m_{\mu\mu} \leq 110\,{\rm GeV}\,,
\end{equation}
on the transverse mass of the virtual W and the invariant mass of the virtual Z. These are needed to suppress non-resonant contributions to the production of the $4$ leptons. Standard acceptance cuts on the charged leptons are also applied. The unweighted events obtained with {\sc MadGraph} are further processed to compute the kinematical variables in eq.~(\ref{feat}) after neutrino reconstruction, as detailed at the beginning of this section. 

The {\sc MadGraph} LO generator is slightly more accurate than the Toy one. It contains all the ZW helicity amplitudes and no high-energy approximation. Furthermore, it describes non-resonant topologies and off-shell vector bosons production, which affects the reconstruction of the neutrino and in turn the reconstruction of the Z and W decay variables~\cite{Panico:2017frx}. Nevertheless on single-variable distributions the Toy Monte Carlo and the LO one agree reasonably well, at around $10\%$.

The third and most refined generator is {\sc MadGraph} at NLO in QCD, interfaced with {\sc{Pythia}}~8.244~\cite{Sjostrand:2006za,Sjostrand:2007gs} for QCD parton showering. The complete $2\to4$ process is generated like at LO, but no cuts could be applied at generation level apart from default acceptance cuts on the leptons and the lower cut on $m_{\mu\mu}$ in eq.~(\ref{eq:cuts_inv_mass}). At NLO, the cut on $p_{T,{\rm{Z}}}$ needs to be replaced with the cut $p_{T,{\rm{V}}}>300$~GeV, with $p_{T,{\rm{V}}}={\rm{min}}[p_{T,{\rm{Z}}},p_{T,{\rm{W}}}]$. This cut suppresses soft or collinear vector boson emission processes, which are insensitive to the EFT. In order to populate the $p_{T,{\rm{V}}}>300$~GeV tail of the distribution with sufficient statistics, events were generated with a bias. The bias function was equal to one for $p_{T,{\rm{V}}}$ above $290$~GeV, and equal to $(p_{T,{\rm{V}}}/290\,{\rm{GeV}})^4$ below. The momenta of the charged leptons and the transverse momentum of the neutrino in the generated events were read with {\sc MadAnalysis}~\cite{Conte:2012fm} and the kinematical variables in eq.~(\ref{feat}) reconstructed like at LO. The cut $p_{T,{\rm{V}}}>300$~GeV and the remaining cuts in eq.~(\ref{eq:cuts_inv_mass}) were imposed on the reconstructed events. The total cut efficiency on the Monte Carlo data, thanks to the bias, was large enough (around $17\%$) to allow for an accurate prediction of the cross section and for the generation of large enough event samples. 

Even if ours is an electroweak process, it is known that NLO QCD corrections can in principle affect significantly the sensitivity to the EFT operators. Relevant effects are related with the tree-level zero~\cite{Baur:1994ia} in the transverse amplitude, which is lifted at NLO, and with the appearance of same-helicity transverse high-energy amplitudes due to real NLO radiation~\cite{Dixon:1993xd}. All these effects are properly modeled by the {\sc MadGraph} NLO generator.

\section{Optimality on Toy data}\label{sec:IR}

Our goal is to reconstruct the EFT-over-SM cross section ratio $r(x,c)$ as accurately as possible using the methods introduced in Section~\ref{sec:teaching}. Since $r$ is known analytically for the Toy problem, a simple qualitative assessment of the performances could be obtained by a point-by-point comparison (see Figures~\ref{fig:scatter} and~\ref{fig:scatter2}) of $r(x,c)$ with its approximation $\widehat{r}(x,c)$ provided by the trained Neural Network. However a point-by-point comparison is not quantitatively relevant, since the level of accuracy that is needed for $\widehat{r}(x,c)$ can be vastly different in different regions of the phase-space, depending on the volume of expected data and on the discriminating power of each region (i.e.~on how much $r$ is different from one). 

The final aim of the entire construction is to obtain an accurate modeling of the extended log-likelihood ratio in eq.~(\ref{llr}), to be eventually employed in the actual statistical analysis. A quantitative measure of the $r$ reconstruction performances is thus best defined in terms of the performances of the final analysis that employs $\widehat{r}$, instead of $r$, in the likelihood ratio. Among all possible statistical analyses that could be carried out, frequentist tests to the EFT hypothesis $H_0(c)$ (regarded as a simple hypothesis for each given value of $c$), against the SM one, $H_1$, are considered for the illustration of the performances.

Four alternative test statistic variables are employed. One is the standard Poisson binned likelihood ratio (see below). The others are unbinned and take the form
\beq\label{tv}
t_c({\mathcal{D}})={\rm{N}}({\rm{X}}|H_{0})-{\rm{N}}({\rm{X}}|H_{1})
- \sum\limits_{i=1}^{\mathcal{N}}\tau_c(x_i)\,,
\eeq
where $\tau_c(x)$ is either equal to the exact $\log[{r(x,c)}]$ or to $\log[{\widehat{r}(x,c)}]$, as reconstructed either with the Standard Classifier or with the Quadratic Classifier described in Section~\ref{SC} and~\ref{PaC}, respectively. In each case the probability distributions of $t$ in the two hypotheses are computed with toy experiments (or with the simpler strategy of Section~\ref{sec:tsdNLO}), and used to estimate the expected (median) exclusion reach on $c$ at $95\%$ Confidence Level if the SM hypothesis is true. In formulas, the $95\%$ reaches ($c_{2\sigma}$) we quote in what follows are solutions to the implicit equation
\beq\label{c95}
p(t_{\rm{med}}(c_{2\sigma});c_{2\sigma})=0.05,\,\;{\rm{with}}\;\;\;t_{\rm{med}}(c)={\rm{Median}}\left[t_c({\mathcal{D}})|H_1\right]\,,
\eeq
where the $p$-value is defined as 
\beq\label{pv}
p(t_c;c)=\int_{t_c}^\infty \hspace{-5pt}dt^\prime_c\, {\rm{pdf}}(t^\prime_c | H_0(c))\,.
\eeq
The two Wilson coefficients $c=G^{(3)}_{\varphi q}$ and $c=G_W$ are considered separately. Therefore the results that follow are single-operator expected exclusion reaches. 

\bigskip

Summarizing, the four methodologies we employ are

\begin{itemize}
\item[i)]{\it Matrix Element (ME)}

In this case we set $\tau_c(x)=\log[{r(x,c)}]$ in eq.~(\ref{tv}), with $r$ computed analytically using eq.~(\ref{acs}). Therefore $t$ coincides with the log-likelihood ratio $\lambda$ in eq.~(\ref{llr}), which in turn is the optimal discriminant between $H_0$ and $H_1$ due to the Neyman--Pearson lemma~\cite{Neyman:1933wgr}. Namely, a straightforward application of the lemma guarantees that by employing $t=\lambda$ as test statistic we will obtain the optimal (smallest) $c_{2\sigma}$ reach, better than the one we could have obtained using any other variable. The Matrix Element Method is thus optimal in this case, and the optimality of the other methods can be assessed by comparing their $c_{2\sigma}$ reach with the one of the Matrix Element.

\item[ii)]{\it Standard Classifier (SC)}

The second method consists in setting $\tau_c(x)=\log[{\widehat{r}(x,c)}]$ in eq.~(\ref{tv}), with $\widehat{r}$ reconstructed by the Standard Classifier as in Section~\ref{SC}. Notice that a separate training is needed to reconstruct $\widehat{r}(x,c)$ for each value of the Wilson Coefficient. Therefore computing $c_{2\sigma}$, as defined in eq.~(\ref{c95}), requires scanning over $c$, performing first the Neural Network training and next the calculation of the distributions of $t$ by toy experiments. For the Quadratic Classifier (and for the Matrix Element Method), the first step is not needed. The details on the Neural Network architecture and training, and of its optimization, will be discussed in Section~\ref{sec:IAV}.

\item[iii)]{\it Quadratic Classifier (QC)}

The third approach is to employ $\widehat{r}(x,c)$ as reconstructed by the Quadratic Classifier of Section~\ref{PaC}. Implementation details are again postponed to Section~\ref{sec:IAV}, however it is worth anticipating that the key for a successful reconstruction is to train using values for the Wilson coefficients that are significantly larger than the actual reach. Specifically, we used
\begin{eqnarray}\label{tranv}
G^{(3)}_{\varphi q} &\ :\quad& \{\pm50, \pm20, \pm5\} \times 10^{-2}\;{\rm TeV}^{-2}\,, \nonumber\\
\rule{0pt}{1.25em}G_W &\ :\quad& \{\pm20, \pm10, \pm5\} \times 10^{-2}\;{\rm TeV}^{-2}\,.
\end{eqnarray}
These values have been selected as those that induce order one departures from the SM cross section in the low, medium and high regions of the $p_{T,{\rm{Z}}}$ distribution. If willing to compute cross-section in each $p_{T,{\rm{Z}}}$ region by quadratic interpolation, using the values selected with this criterion can be shown to maximize the accuracy on the reconstruction of the linear term, while still allowing for a good determination of the quadratic term. We expect this choice to be optimal for the Quadratic Classifier training as well. Also notice that the total number of training Monte Carlo events is the same one ($6$M, see Section~\ref{sec:IAV}) employed for each of the separate trainings performed on the Standard Classifier.

\item[iv)]{\it Binned Analysis (BA)}

Finally, in order to quantify the potential gain of the unbinned strategy, we also perform a binned analysis. The test statistic in this case is provided by the sum over the bins of the log-ratio of the SM over EFT Poisson likelihoods, with the expected countings as a function of the Wilson coefficients computed from Monte Carlo simulations. The test statistic distributions, and in turn the reach by eq.~(\ref{c95}), are computed with toy experiments like for the other methods and no asymptotic formulas are employed.

For both $G^{(3)}_{\varphi q}$ and $G_W$ we considered $3$ bins in $p_{T,{\rm{Z}}}$, with the following boundaries
\begin{equation}
p_{T,{\rm{Z}}} [{\rm GeV}]\ :\quad \{300, 500, 750, 1200\}\;{\rm GeV}\,.
\end{equation}
The $p_{T,{\rm{Z}}}$ variable is an extremely important discriminant because it is sensitive to the energy growth induced by the EFT. The three bins are selected based on the studies in Refs.~\cite{Panico:2017frx,Franceschini:2017xkh}, and a narrower binning has been checked not to improve the sensitivity significantly. A cut $\cos \Theta < 0.5$ is imposed in the analysis targeting $G^{(3)}_{\varphi q}$, because this helps~\cite{Franceschini:2017xkh} in isolating the longitudinal helicity channel thanks to the amplitude zero in the transverse SM amplitudes. For ${\cal O}_W$, no $\cos \Theta$ cut is performed, and each $p_{T,{\rm{Z}}}$ is split in two bins, for $\cos 2 \,\varphi_W$ larger and smaller than zero. This is sufficient to partially capture the leading EFT/SM interference term as discussed in Ref.~\cite{Panico:2017frx}.

Most likely the binned analysis could be improved by considering more (and/or better) variables and a narrower binning. However it should be noticed that the simple strategies described above already result from an optimization, targeted to the specific operators at hand, and that the reach we obtain is consistent with the sensitivity projections available in the literature.

\end{itemize}

\subsection{Results}

\begin{figure}
\centering
\includegraphics[width=.495\textwidth]{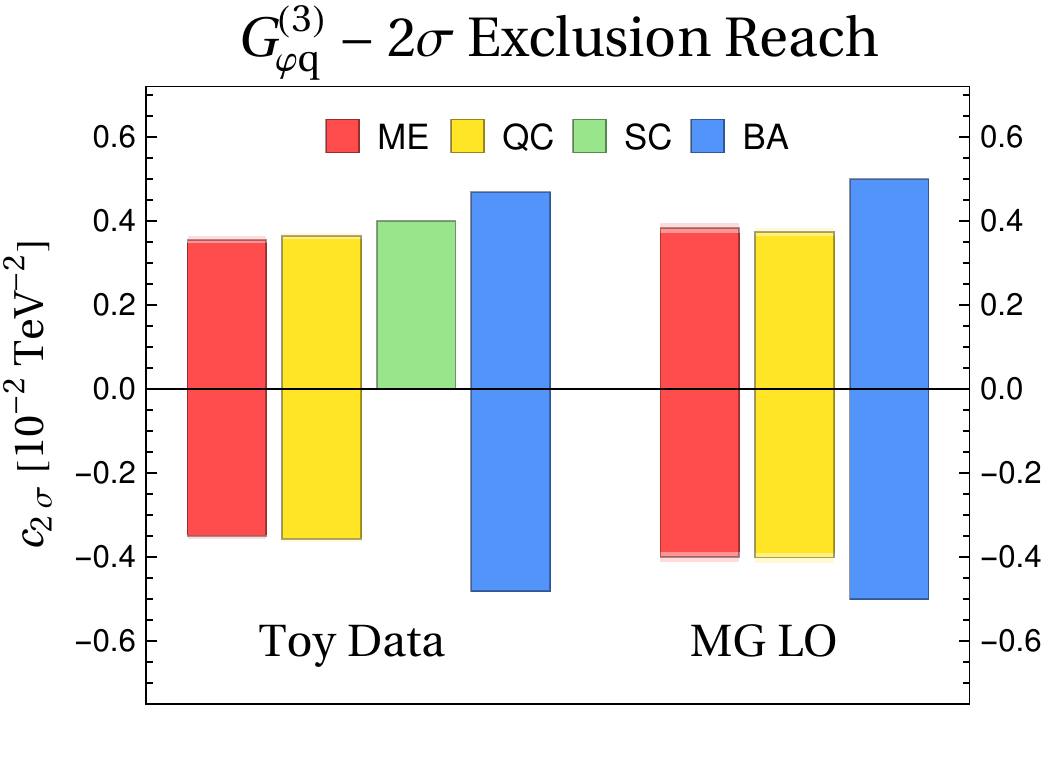}
\hfill
\includegraphics[width=.47\textwidth]{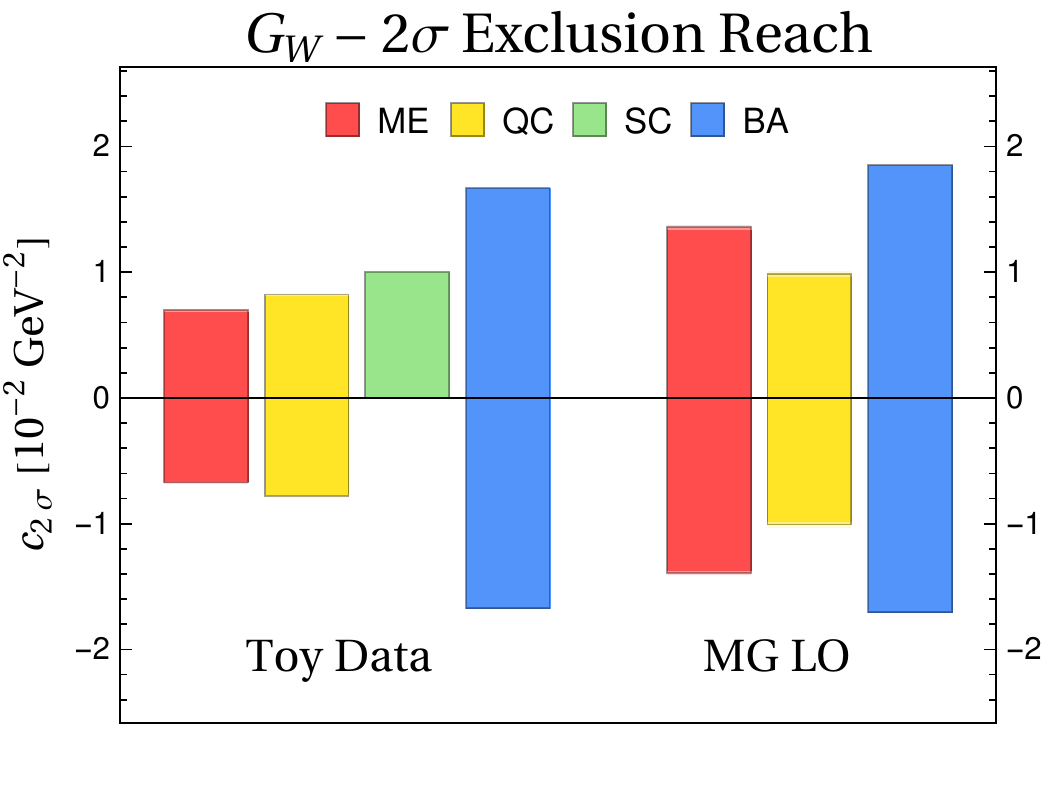}
\caption{Expected exclusion reach on $G^{(3)}_{\varphi q}$ (left) and on $G_W$ (right). The results are also reported in Table~\ref{tab:bounds}. Light-color stacked bars represent the errors.} \label{fig:fits_gphi}
\end{figure}

The results of the four methods are shown in Figure~\ref{fig:fits_gphi} (see also Table~\ref{tab:bounds}), together with the ones obtained with the {\sc MadGraph}~LO description of the ZW process, to be discussed in Section~\ref{MGLO}. The $2\sigma$ sensitivities reported in the figure are obtained by interpolating the median $p$-value as a function of the Wilson Coefficient $c$, in the vicinity of the reach, and computing $c_{2\sigma}$ by solving eq.~(\ref{c95}). Further details on this procedure, and the associated error, are given in Section~\ref{sec:tsdNLO}. 

The figure reveals a number of interesting aspects. First, by comparing the Matrix Element reach with the one of the Binned Analysis we can quantify the potential gain in sensitivity offered by a multivariate strategy. The improvement is moderate (around $30\%$) for $G^{(3)}_{\varphi q}$, but it is more than a factor of $2$ (of $2.4$)
in the case of the $G_W$ operator coefficient. The different behavior of the two operators was expected on physical grounds, as discussed in details below. The figure also shows that the Quadratic Classifier is nearly optimal. More precisely, the reach is identical to the one of the Matrix Element for $G^{(3)}_{\varphi q}$, and $<20\%$ worse for $G_W$. We will see in Section~\ref{sec:IAV} that the residual gap for $G_W$ can be eliminated with more training points than the ones used to produce Figure~\ref{fig:fits_gphi}. Suboptimal performances are shown in the figure in order to outline more clearly, in Section~\ref{sec:IAV}, that our method is systematically improvable as long as larger and larger Monte Carlo samples are available. 

Finally, we see in the figure that the Standard Classifier is slightly less sensitive than the Quadratic one, but still its performances are not far from optimality. This is reassuring in light of possible applications of Statistical Learning methodologies to different problems, where the dependence of the distribution ratio on the new physics parameters is not known and the Quadratic Classifier approach cannot be adopted. On the other hand, the Standard Classifier method is rather demanding. First, because it requires separate trainings on a grid of values of $c$, out of which the reach should be extracted by interpolation. In turn, this requires a much larger number of training points than the Quadratic Classifier, since at each point of the grid we use as many training points as those the Quadratic Classifier needs in total for its training. Second, because we observed hyperparameters optimization depends  on the specific value of $c$ that is selected for training. Because of these technical difficulties, we only report sensitivity estimates for the positive Wilson coefficients reach. Furthermore these estimates (see Table~\ref{tab:bounds}) are based on the $p$-value obtained at a given point of the $c$ grid without interpolation. For the same reason, we did not try to apply the Standard Classifier methodology to the LO and to the NLO data and we focus on the Quadratic Classifier in what follows.

Let us discuss now the physical origin of the different behaviors observed for the ${\cal O}^{(3)}_{\varphi q}$ and for the ${\cal O}_W$ operator. The point is that the new physics effects due to ${\cal O}_{\varphi q}^{(3)}$ have very distinctive features which can be easily isolated even with a simple binned analysis with few variables. Indeed ${\cal O}_{\varphi q}^{(3)}$ (see eq.~(\ref{eq:amplitudes})) only contributes to the $00$ polarization amplitude, which is non-vanishing in the SM as well and proportional to $\sin\Theta$. The squared $00$ amplitude thus contributes to the cross section with a sizable interference term, which is peaked in the central scattering region $\cos \Theta \sim 0$. The other helicity channels play the role of background, and are peaked instead in the forward region. They are actually almost zero (at LO) at $\cos \Theta \simeq 0$. Therefore a binned analysis targeting central scattering (this is why we imposed the cut $\cos \Theta < 0.5$) is sufficient to isolate the effects of ${\cal O}_{\varphi q}^{(3)}$ at the interference level and thus to probe $G_{\varphi q}^{(3)}$ accurately. By including the decay variables as in the multivariate analysis we gain sensitivity to new terms in the cross section, namely to the interference between the $00$ and the transverse amplitudes, however these new terms are comparable with those that are probed already in the Binned Analysis and thus they improve the reach only slightly. 

The situation is very different for the ${\cal O}_W$ operator. It contributes to the $++$ and $--$ helicity channels, that are highly suppressed in the SM and set exactly to zero in the Toy version of the problem we are studying here. The $p_{T, {\rm{Z}}}$  (and $\Theta$) distribution depends only at the quartic level on $G_W$, i.e.~through the square of the BSM amplitude, because the interference between different helicity channels cancels out if we integrate the cross section in eq.~(\ref{acs}) over the ZW azimuthal decay angles. Our Binned Analysis is sensitive to the interference term through the binning in $\varphi_W$, however this is not enough to approach the optimal reach because all the other decay variables (and $\Theta$ as well) do possess some discriminating power, from which we can benefit only through a multivariate analysis. More specifically, one can readily see by direct calculation that the dependence on all our kinematical variables of the $G_W$ interference contribution to the differential cross section is different from the SM term. By integrating on any of this variables we partially lose sensitivity to this difference, and this is why the multivariate analysis performs much better than the binned one. 

\subsection[{{\sc{MadGraph}} Leading Order}]{{{\sc{\textbf{MadGraph}}}} Leading Order}\label{MGLO}

The analyses performed for the Toy dataset can be easily replicated for the {\sc MadGraph}~LO Monte Carlo description of the process, obtaining the results shown in Figure~\ref{fig:fits_gphi}. 

The most noticeable difference with what was found with the Toy Monte Carlo is the strong degradation of the Matrix Element reach, and the fact that it gets weaker than the one of the Quadratic Classifier. As usual, the effect is more pronounced for the ${\mathcal{O}}_W$ operator. This is not mathematically inconsistent because the analytic ratio $r(x,c)$ we employ for the Matrix Element test statistic is not equal anymore to the ratio of the true distributions according to which the data are generated. Therefore it is not supposed to give optimal performances. On the other hand the observed degradation is quantitatively surprising for $G_W$, especially in light of the fact that the {\sc MadGraph}~LO Monte Carlo distributions seem quite similar to the ones of the Toy data at a superficial look. The degradation is not due to the high-energy approximation in the ZW production amplitude, indeed the results we are reporting are obtained with the exact tree-level helicity amplitudes, which are employed in eq.~(\ref{acs}) in place of the ones in eq.~(\ref{eq:amplitudes}). It is due to the other approximations we performed in the calculation of the cross section, namely to the assumption that the initial quark is always more energetic than the anti-quark, which allows us to interpret $\Theta$ as the angle between the quark and the Z, and to the one of a perfect reconstruction (up to the ambiguity) of the neutrino momentum. We verified that this is the case by repeating the Matrix Element analysis using the true neutrino momentum and the actual direction of motion of the quark in the Monte Carlo events. In this case the reach on $G_W$ gets closer to the one obtained with the Toy data. 

The degradation of the Matrix Element reach should be contrasted with the relative stability of the Quadratic Classifier method. Notice that the method is applied on the {\sc MadGraph}~LO data in the exact same way as on the Toy data, namely the architecture is the same, as well as the number of training point and the values of the Wilson coefficients in eq.~(\ref{tranv}) used for training. The computational complexity of the distribution ratio reconstruction is thus identical in the two cases, in spite of the fact that the  {\sc MadGraph}~LO Monte Carlo offers a slight more complete (or ``complex'') description of the data. The total computational cost is somewhat higher in the {\sc MadGraph}~LO case, but just because the process of Monte Carlo events generation is in itself more costly. Similar considerations hold at NLO, where the cost of event generation substantially increases. 

\section{The reach at Next-to-Leading Order}\label{sec:NLO}

Including NLO QCD corrections is in general essential for an accurate modeling of the LHC data. Therefore it is imperative to check if and to what extend the findings of the previous section are confirmed with the {\sc MadGraph}~NLO Monte Carlo description of the process, introduced in Section~\ref{mcgen}. As far as the reconstruction of ${\widehat{r}}(x,c)$ is concerned, using {\sc MadGraph}~NLO does not pose any conceptual or technical difficulty, provided of course the (positive and negative) Monte Carlo weights are properly included in the loss function as explained in Section~\ref{sec:teaching}. Computing the distribution of the test statistic variable that we obtain after the reconstruction (or of the one we employ with the Matrix Element method, for which the exact same issue is encountered) is instead slightly more complicated than with the Toy and {\sc MadGraph}~LO data. This point is discussed in the following section, while the illustration of the results is postponed to Section~\ref{resNLO}.

\subsection{Estimating the test statistics distributions}\label{sec:tsdNLO}

As soon as $\tau_c(x)$ is known, either as an analytic function in the case of the Matrix Element or as a (trained) Neural Network in the case of the Quadratic Classifier, the test statistic $t_c({\mathcal{D}})$, as defined in eq.~(\ref{tv}), is fully specified. Namely we can concretely evaluate it on any dataset ${\mathcal{D}}=\{x_i\}$, consisting of ${\mathcal{N}}$ repeated instances of the variable $x$, for each given value of $c$. However in order to perform the hypothesis test, and eventually to estimate the reach $c_{2\sigma}$, we also need to estimate the probability distribution of $t_c({\mathcal{D}})$ under the $H_0$ and under the $H_1$ hypotheses. This is the problematic step at NLO, after which the evaluation of $c_{2\sigma}$ proceeds in the exact same way as for the Toy and for the LO data. Specifically, once we are given with
\beq\label{pdft}
{\rm{pdf}}(t_c | H_0(c))\;\;{\rm{and}}\;\;{\rm{pdf}}(t_c | H_1)\,,
\eeq
we obtain the $p$-value as a function of $t_c$ and $c$ as in eq.~(\ref{pv}) from the former, while from the latter we compute the median value of $t_c$ and in turn 
\beq\label{pmed}
p_{\rm{med}}(c)\equiv p(t_{\rm{med}}(c);c)\,,
\eeq
as a function of $c$. After scanning over $c$ and interpolating $p_{\rm{med}}(c)$ in the vicinity of the reach (actually we interpolate the logarithm of $p_{\rm{med}}(c)$, using three points in $c$ and quadratic interpolation), we can solve the equation $p_{\rm{med}}(c_{2\sigma})=0.05$ and obtain the reach as defined in eq.~(\ref{c95}). Given the error on $p_{\rm{med}}(c)$ at the three points used for the interpolation, the error on the estimate of $c_{2\sigma}$ is obtained by error propagation. 

It is conceptually trivial (but numerically demanding) to estimate the distributions if artificial instances of the dataset ${\mathcal{D}}$ (aka ``toy'' datasets) are available. In this case one can simply evaluate $t_c({\mathcal{D}})$ on many toy datasets following the $H_0(c)$ and the $H_1$ hypotheses and estimate the distributions. More precisely, one just needs the empirical cumulative in $H_0(c)$ and the median of $t_c$ in $H_1$. Toy datasets are readily obtained from unweighted Monte Carlo samples by throwing ${\cal{N}}$ random instances of $x$ from the sample, with ${\cal{N}}$ itself thrown Poissonianly around the total expected number of events. This is impossible at NLO because the events are necessarily weighted, therefore they are not a sampling of the underlying distribution of the variable $x$. As emphasized in Section~\ref{sec:teaching}, NLO Monte Carlo data can only be used to compute expectation values of observables $O(x)$ as in eq.~(\ref{LS}). For instance we can compute the cross section in any region of the X space, and the mean or the higher order moments of the variable of interest, $\tau_c(x)$.

This suggests two options to estimate the distributions of the test statistic at NLO. The first one is to compute the distribution of $\tau_c(x)$ by means of a (weighted) histogram with many and very narrow bins. By knowing the cross section of each bin in $\tau_c$, we know how many events are expected to fall in that bin and generate toy datasets for $\tau_c$ accordingly. This procedure is quite demanding, and it relies on a careful choice of the $\tau_c$ binning, which can only be performed on a case-by-case basis. It is still useful to validate the strategy we actually adopt, described below.

\begin{table}[t]
\centering
{
\begin{tabular}{c|c|ccc}
\multicolumn{1}{c}{} & & Toy Data & LO & NLO\\
\midrule
\multirow{4}{*}{\rule{0pt}{2.5em}$G^{(3)}_{\varphi q}$}  & \rule{0pt}{1.25em}ME & $[-0.350(6), 0.356(8)]$ & $[-0.399(13), 0.384(12)]$ & $[-0.55(4), 0.464(14)]$\\
& \rule{0pt}{1.5em}SC & $\gtrsim 0.4\,(p = 0.077(5))$ & --- & ---\\
& \rule{0pt}{1.5em}QC & $[-0.357(6), 0.365(8)]$ & $[-0.401(12), 0.374(10)]$ & $[-0.426(22), 0.401(21)]$\\
& \rule[-.75em]{0pt}{2.25em}BA & $[-0.48, 0.47]$ & $[-0.50, 0.50]$ & $[-0.58, 0.55]$\\
\midrule
\multirow{4}{*}{\rule{0pt}{2.5em}$G_{W}$} & \rule{0pt}{1.25em}ME & $[-0.673(14), 0.697(11)]$ & $[-1.390(21), 1.357(22)]$ & $[-1.51(7), 1.93(14)]$\\
& \rule{0pt}{1.5em}SC & $\lesssim 1\,(p=0.038(3))$ & --- & ---\\
& \rule{0pt}{1.5em}QC & $[-0.781(13),0.822(13)]$ & $[-1.007(27),0.987(26)]$ & $[-0.99(4), 1.08(12)]$\\
& \rule[-.75em]{0pt}{2.25em}BA & $[-1.67, 1.67]$ & $[-1.70, 1.85]$ & $[-1.63, 1.98]$
\end{tabular}
}
\caption{Bounds on the $G^{(3)}_{\varphi q}$ and $G_W$ coefficients obtained for the Toy, LO and NLO datasets.
The rows correspond to the Matrix Element (ME), Standard Classifier (SC), Quadratic Classifier (QC) and Binned Analysis (BA)
approach. Notice that the errors on the Binned Analysis bounds are negligible. The results are given in $10^{-2}\,{\rm TeV}^{-2}$ units.}\label{tab:bounds}
\end{table}

The second option is to approximate the distribution of $t_c$ in a ``nearly Gaussian'' form, based on the Central Limit theorem. Namely we notice that $t_c$ is in a trivial linear relation (see eq.~(\ref{tv})) with the variable
\beq
{\mathcal{T}}_c(\mathcal{D})\equiv\frac1{\rm{N}}\sum\limits_{i=1}^{\cal{N}}\tau_c(x_i)\,,
\eeq
where $\cal{N}$ is Poisson-distributed with expected ${\rm{N}}$, with ${\rm{N}}={\rm{N}}({\rm{X}}|H)$  and $H=H_0$ or $H=H_1$. The $x_i$'s are independent and sampled according to \mbox{pdf$(x|H)$}. The cumulant-generating function of ${\mathcal{T}}_c$ (which is a so-called ``compound'' Poisson variable~\cite{CPo}) is readily computed
\beq
{\rm{K}}_{{\mathcal{T}}_c}(\xi)\equiv\log\left\{{\rm{E}}\left[\left.
e^{\xi {\mathcal{T}}_c}\right|\,H
\right]\right\}= {\rm{N}} \, {\rm{E}}\left[\left.
e^{\frac\xi{N} \tau_c}\right|\,H
\right]
-{\rm{N}}\,,
\eeq
by first taking the expectation on the $x_i$'s conditional to ${\cal{N}}$ and next averaging over the Poisson distribution of  ${\cal{N}}$. Therefore the cumulants of ${\mathcal{T}}_c$,
\beq\label{cumu}
{\kappa}^n_{{\mathcal{T}}_c}\equiv\left.\frac{d^n {\rm{K}}_{{\mathcal{T}}_c}(\xi)}{d\xi^n}\right|_{\xi=0}={{\rm{N}}^{1-n}}{\rm{E}}\left[\left.
\tau_c^n\right|\,H
\right]\,,
\eeq
are increasingly suppressed with ${\rm{N}}$ for larger and larger $n>1$. Since ${\rm{N}}$ is of the order of several thousands in our case, neglecting all cumulants apart from the first and the second one, i.e. adopting a Gaussian distribution for ${\mathcal{T}}_c$, might be a good approximation.

Actually it turns out that in order to model properly the $5\%$ tail of the distribution, which we need to probe for the exclusion limit, non-Gaussianity effects can be relevant. These are included by using a skew-normal distribution for ${\mathcal{T}}_c$, which contains one more adjustable parameter than the Gaussian to model the skewness. The mean, standard deviation and skewness of ${\mathcal{T}}_c$ are immediately obtained from eq.~(\ref{cumu})
\beq
\mu({{\mathcal{T}}_c})=\langle\tau_c\rangle\,,\;\;\;\;\;\sigma({{\mathcal{T}}_c})=\frac1{\sqrt{\rm{N}}}\sqrt{\langle\tau_c^2\rangle}\;\;\;\;\;\mu_3({{\mathcal{T}}_c})=\frac1{\sqrt{\rm{N}}}\frac{\langle\tau_c^3\rangle}{{\langle\tau_c^2\rangle}^{3/2}}\,,
\eeq
where $\langle\cdot\rangle$ is used to denote expectation for brevity. By computing the expectation values of $\tau_c$, $\tau_c^2$ and $\tau_c^3$ using the Monte Carlo data, we thus find the parameters of the skew-normal distribution for ${\mathcal{T}}_c$ and in turn the distribution of $t_c$. We finally obtain the median p-value from the definition in eq.~(\ref{pmed}). The errors on the expectation values of $\tau_c$ are estimated from the fluctuations in the means on subsets of the entire Monte Carlo sample. These errors are propagated to the $p$-value and eventually to the $c_{2\sigma}$ estimated reach as previously explained. Accurate results (see Table~\ref{tab:bounds}) are obtained with relatively small Monte Carlo samples. Namely, $500$k event were used at NLO, $1$M at LO and $3$M for the Toy data.

\begin{figure}
	\centering
	\includegraphics[width=.495\textwidth]{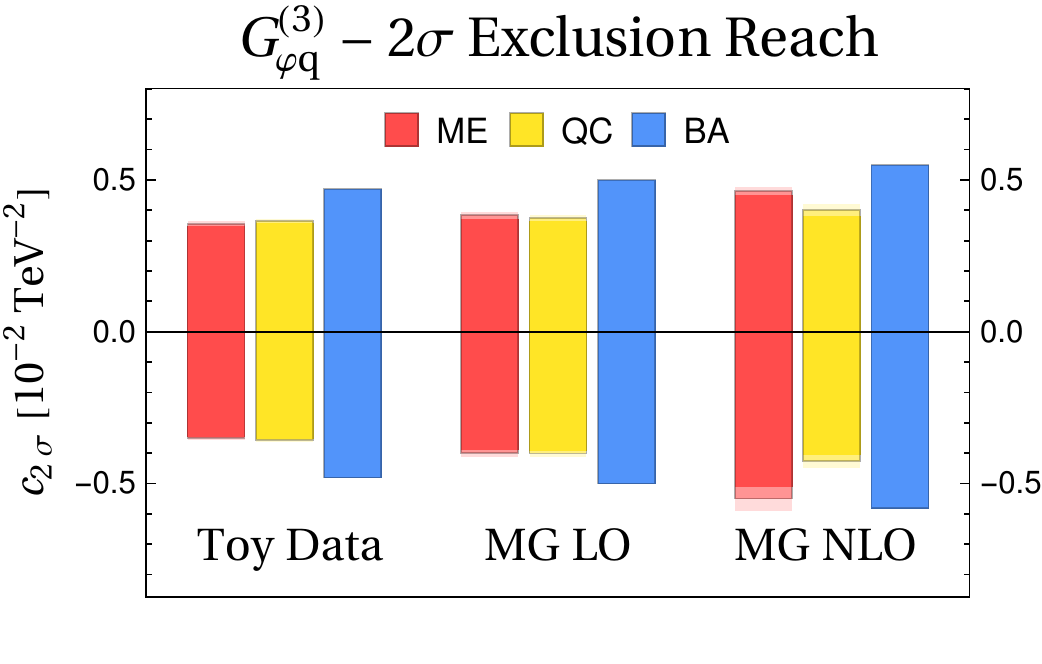}
	\hfill
	\includegraphics[width=.47\textwidth]{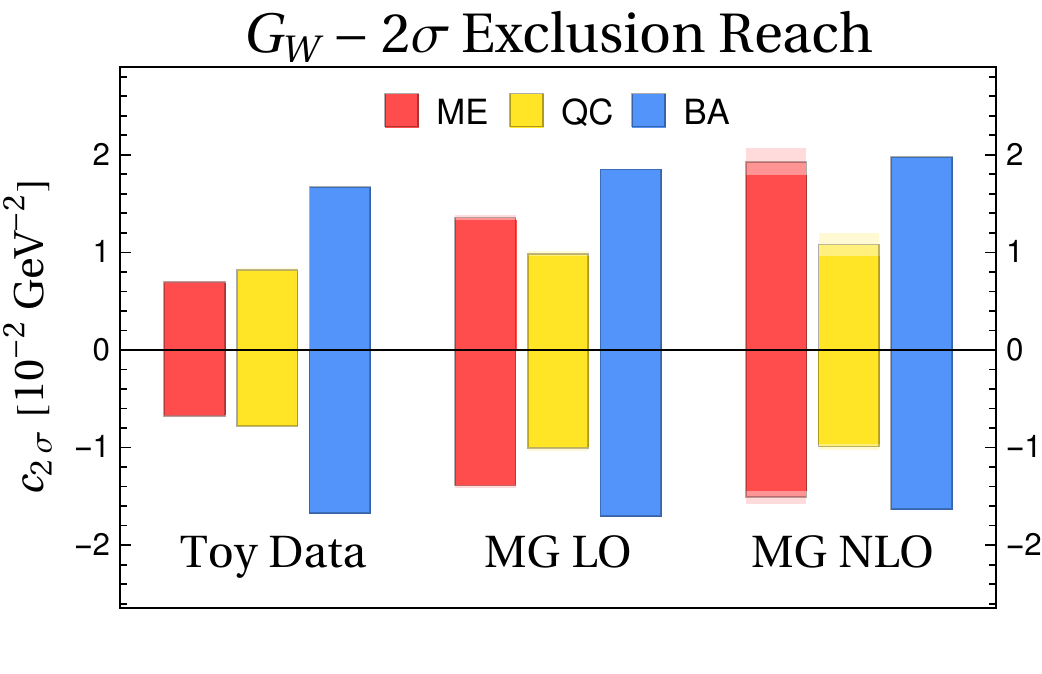}
	\caption{Expected exclusion reach on $G^{(3)}_{\varphi q}$ (left) and on $G_W$ (right) with the various methodologies described in the text. The results are also reported in Table~\ref{tab:bounds}.} \label{fig:fits_gphiNLO}
\end{figure}

We cross-checked the above procedure in multiple ways. First, it reproduces within errors the LO and Toy $p$-values obtained with the toy experiments. Second, we validated it against the approach based on $\tau_c$ binning on NLO data, as previously mentioned. We also verified that including the skewness changes the results only slightly, with respect to those obtained in the Gaussian limit. Further improving the modeling of the non-Gaussiantiy with more complex distributions than the skew-normal, with more adjustable parameters in order to fit higher order moments of ${{\mathcal{T}}_c}$, is therefore not expected to affect the results.

\subsection{Results}\label{resNLO}

Our results with the {\sc MadGraph}~NLO Monte Carlo are reported in Figure~\ref{fig:fits_gphiNLO} and in Table~\ref{tab:bounds}. They essentially confirm the trend we already observed in the transition from the Toy to the {\sc MadGraph}~LO data. The Matrix Element keeps losing sensitivity because the analytic distribution ratio is now even more faraway from the actual distribution ratio since it does not include NLO QCD effects. The reach of the Binned Analysis deteriorates less, so that it becomes comparable to the one of the Matrix Element. The Quadratic Classifier reach is remarkably stable. Actually it slightly improves with respect to the LO one for $G_W$. This is probably due to the appearance of same-helicity SM transverse amplitudes (see Section~\ref{mcgen}) and of the corresponding interference term for the ${\mathcal{O}}_W$ operators.

Notice few minor differences in the implementation of the Quadratic Classifier and of the Binned Analysis at NLO. The Quadratic Classifier now also employs the variable $p_{T,{\rm{ZW}}}$, as discussed in Section~\ref{sec:WZ}. The Binned Analysis for $G^{(3)}_{\varphi q}$ employs $p_{T,{\rm{ZW}}}$ as well, through a cut $p_{T,{\rm{ZW}}}/p_{T,{\rm{V}}}<0.5$. This improves the reach~\cite{Franceschini:2017xkh} because it helps recovering (partially) the background suppression due to the zero of the transverse amplitudes in the central region.

\section{Neural Network implementation and validation}\label{sec:IAV}

\begin{figure}
	\centering
	\includegraphics[width=0.48\linewidth]{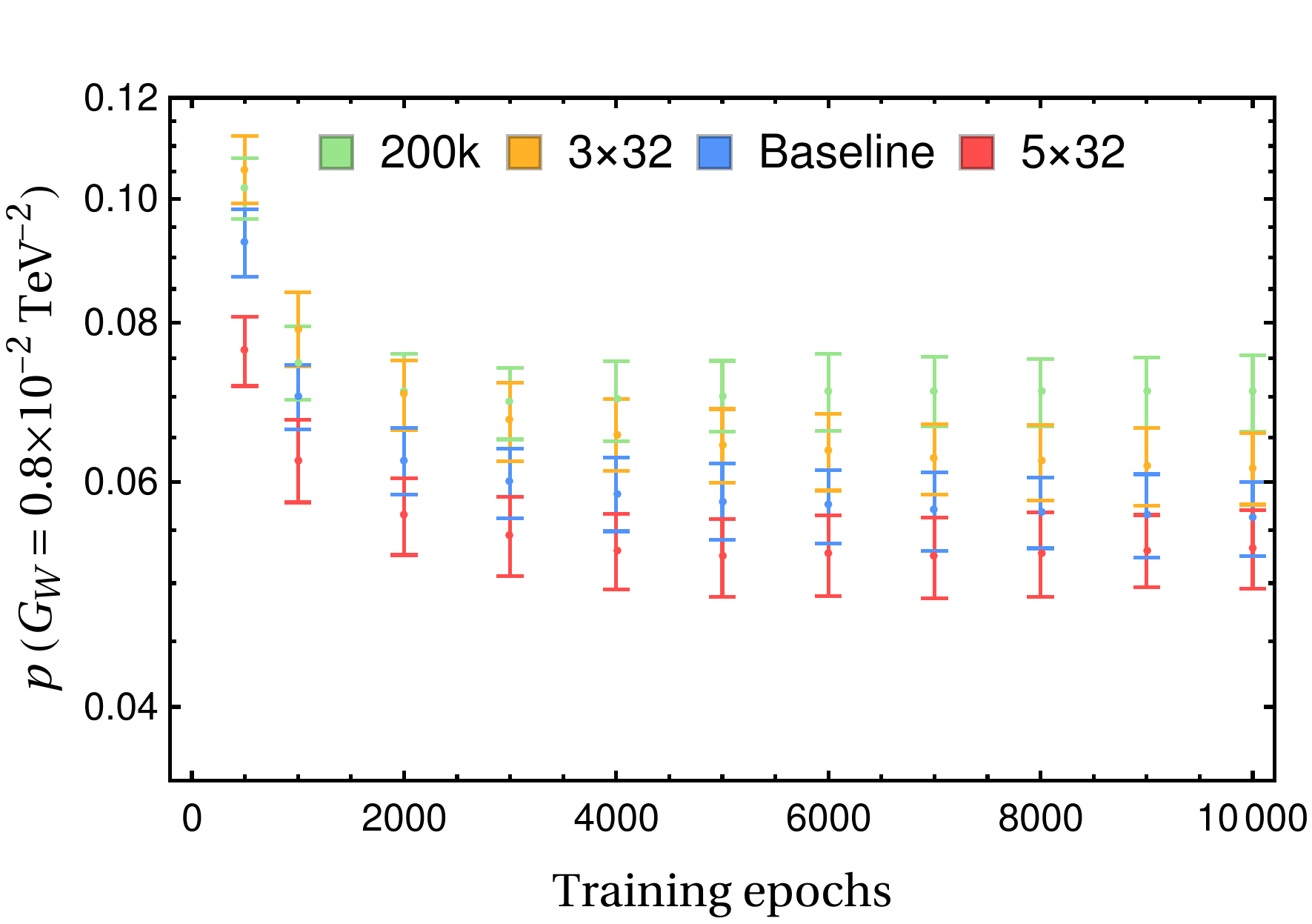}\hfill
	\includegraphics[width=0.48\linewidth]{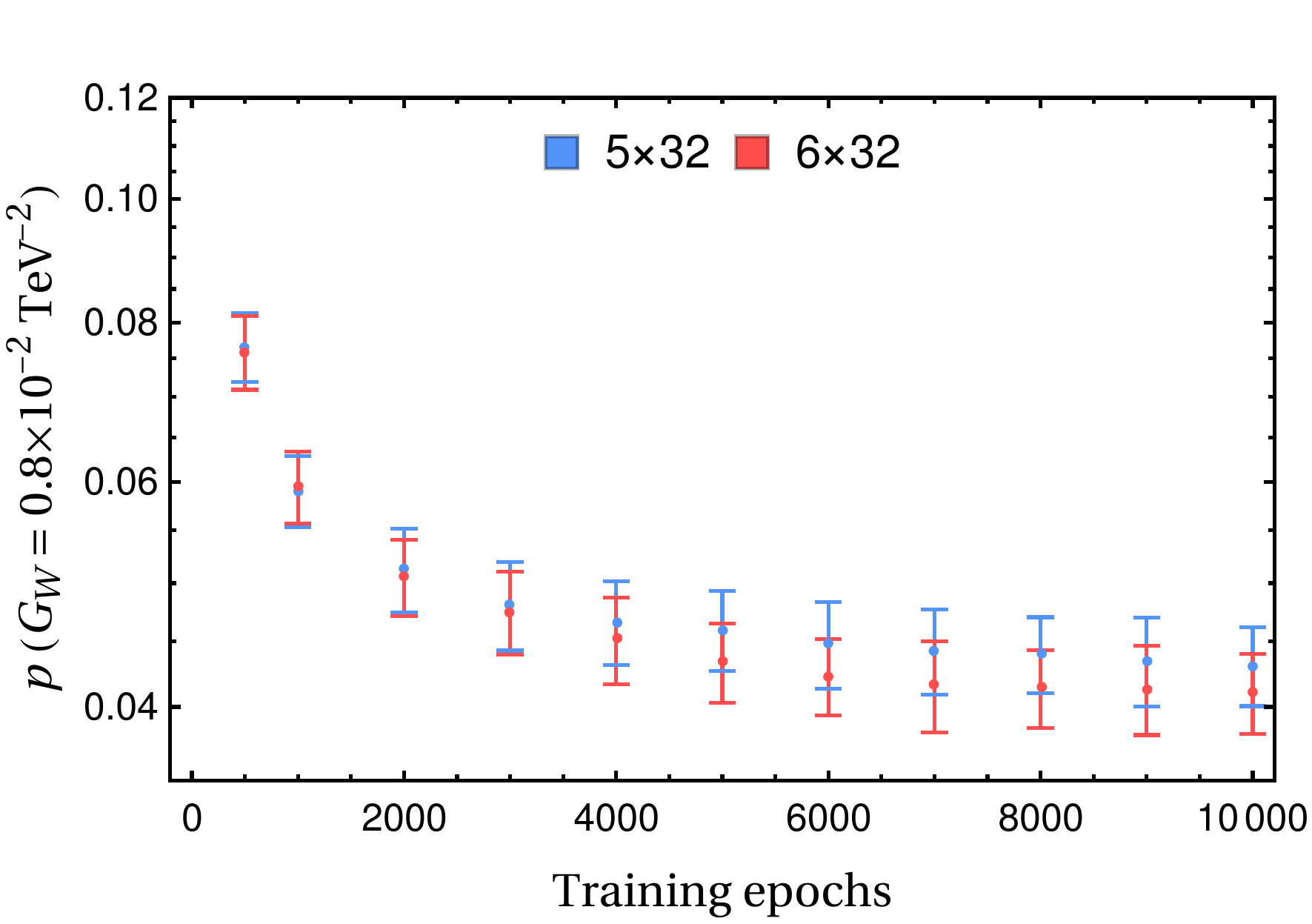}
	\caption{Evolution of the $p$-value for different architectures and training sample sizes. On the left plot we compare the baseline setup with the baseline architecture Network  trained with $200$k points per value of $c$ (for a total of $2.4$M points), and with the baseline number of training points  ($500$k, times $12$) on architectures with one less (``$3\times32$'') and one more (``$5\times32$'') hidden layer. On the right plot, a similar analysis is performed, but with $3$M points per value of $c$.}
	\label{fig:pvalueevolution}
\end{figure}

The strategies described in Section~\ref{sec:teaching} were implemented in {\tt{Pytorch}}~\cite{NEURIPS2019_9015} and run on NVIDIA GeForce GTX~1070 graphics card. Fully connected feedforward deep Neural Networks were employed, acting on the features vector
\beq
x=\{ s,\,\Theta,\, \theta_W,\,\theta_Z,\, p_{T,{\rm{ZW}}},\, p_{T,{\rm{Z}}},\,\sin \varphi_W,\,\cos \varphi_W,\,\sin \varphi_Z,\,\cos \varphi_Z  \}\,,
\eeq
for a total of $10$ features. Each feature is standardized with a linear transformation to have zero mean and unit variance on the training sample. For the Quadratic Classifier training, the Wilson coefficient employed in the parametrization~(\ref{PC}) were scaled to have unit variance on the training sample. Employing the redundant variables (i.e., $ p_{T,{\rm{Z}}}$, and the cosines and sines of $\varphi_{W,Z}$) is helpful for the performances, especially the angular ones,
which enforce the periodicity of the azimuthal angular variables. The ``baseline'' results presented in Figures~\ref{fig:fits_gphi}, \ref{fig:fits_gphiNLO} and in Table~\ref{tab:bounds} were all obtained with the features vector above and employing a total of $6$ million training  Monte Carlo points for each of the two Wilson coefficients. Training was always performed with a single batch (which was found to perform better in all cases), even if in practice the gradients calculation was split in mini-batches of $100$k points in order to avoid saturating the memory of the GPU. Apart from these common aspects, the optimization of the Neural Network design and of the training strategy is rather different for the Quadratic and for the Standard Classifier methods. They are thus discussed separately in what follows.

\subsection{The Quadratic Classifier}\label{PCO}

For the Quadratic Classifier, best performances were obtained with ReLU activation functions and with the Adam~{\tt{Pytorch}} optimizer. The initial learning rate (set to $10^{-3}$) does not strongly affect the performances. Other attempts, with Sigmoid activation and/or with SGD optimizer, produced longer execution time and worse performances. The baseline architecture for the two Neural Networks ${\rm{n}}_{\alpha}$ and ${\rm{n}}_{\beta}$ in eq.~(\ref{PC}) consists of $4$ hidden layers with $32$ neurons, namely the architecture $\{10,32,32,32,32,1\}$, including the input and the output layers. Weight Clipping was implemented as a bound on the $L_1$ norm of the weights in each layer, but found not to play a significant role. The total training time, for $10^{4}$ training epochs, is around $5$ hours for the baseline architecture and with the baseline number ($6$ million) of training points.

The Neural Network architecture was selected based on plots like those in Figure~\ref{fig:pvalueevolution}. The left panel shows the evolution with the number of training epochs of the median $p$-value (see eq.~(\ref{pmed})) on Toy data for $c=G_W=0.8\times10^{-2}\,{\rm TeV}^{-2}$, with the baseline and with larger and smaller Networks. We see that adding or removing one hidden layer to the baseline architecture does not change the performances significantly. The plot also shows that $10^{4}$ epochs are sufficient for the convergence and that no overfitting occurs. The degradation of the performances with less training point is also illustrated in the plot. Of course, the $p$-value is evaluated using independent Monte Carlo samples, not employed for training. The errors on the $p$-value are estimated from the error on the skew-normal distribution parameters as explained in Section~\ref{sec:tsdNLO}. In the baseline configuration we used $500$k EFT Monte Carlo training points for each of the $6$ values of $G_W$ in eq.~(\ref{tranv}), plus $500$k for each associated SM sample. Each sample consists instead of $3$M points in the extended configuration employed on the right panel of Figure~\ref{fig:pvalueevolution}, for a total of $36$M. The same value of $G_W=0.8\times10^{-2}\,{\rm TeV}^{-2}$ is employed. The baseline architecture becomes insufficient, and best results are obtained with the $6$ hidden layers of $32$ neurons each. 

\begin{figure}
	\centering
	\includegraphics[width=0.45\linewidth]{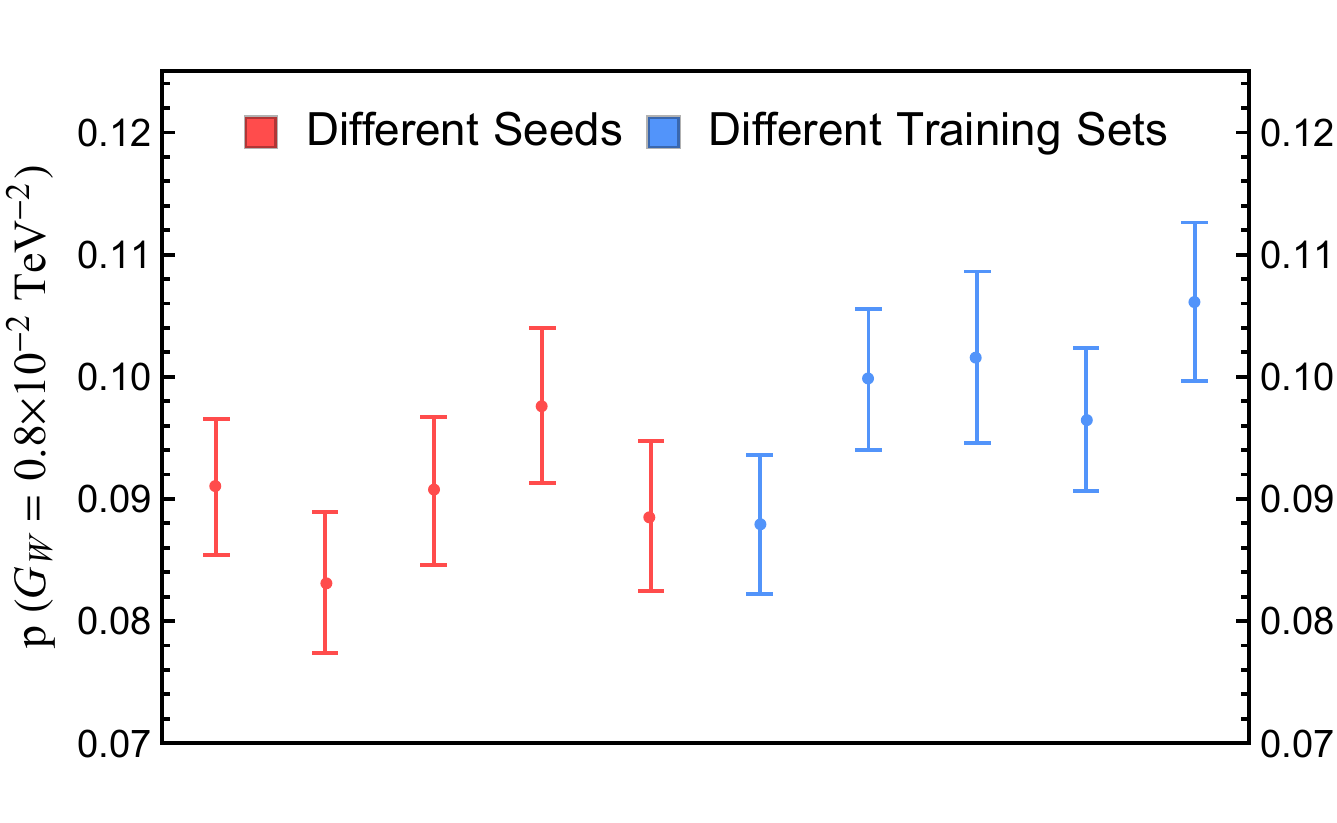}
	\caption{Results of $5$ different trainings of the same architecture (Baseline architecture trained with $2.4$M points) using: the same training data but different initialization seeds (red points) and the same initialization but different training data samples (blue points).}
	\label{fig:trainingstability}
\end{figure}

The figure also demonstrates that the method is systematically improvable towards optimality. The value of $G_W$ considered in the figure was not within the $95\%$ CL reach with the baseline setup, while it becomes visible with the extended configuration. All the reaches reported in Table~\ref{tab:bounds} would expectedly improve with the extended configuration. The $G_W$ reach on Toy data becomes \mbox{$[-0.732(9), 0.764(14)]\,10^{-2}\,$TeV$^{-2}$}, which is now only less than $10\%$ worse than the optimal Matrix Element reach. Training takes around $30$ hours with the extended configuration, while generating and processing the required training points with {\sc{MadGraph}}~NLO (which is the most demanding generator) would take around $10$ days on a $32$-cores workstation. We could thus try to improve also the NLO reach even with limited computing resources.

For the reproducibility of our results we also study how the performances depend on the Neural Network initialization and on the statistical fluctuations of the Monte Carlo training sample. This analysis is performed in a reduced setup, with a total of $2.4$ million training point, and for $G_W=0.8\times10^{-2}\,{\rm TeV}^{-2}$. We see in Figure~\ref{fig:trainingstability} that the $p$-value fluctuates by varying the random seed used for training at a level comparable with the error on its determination. Similar results are observed by employing different independent Monte Carlo training samples. Notice that these fluctuations should not be interpreted as additional contributions to the error on the $p$-value. Each individual Neural Network obtained from each individual training defines a valid test statistic variable, on which we are allowed to base our statistical analysis. Since the fluctuations are comparable to the $p$-value estimate errors, our sensitivity projections were obtained by randomly selecting one of the seed/training set configuration. 

\begin{figure}
	\centering
	\includegraphics[width=0.45\linewidth]{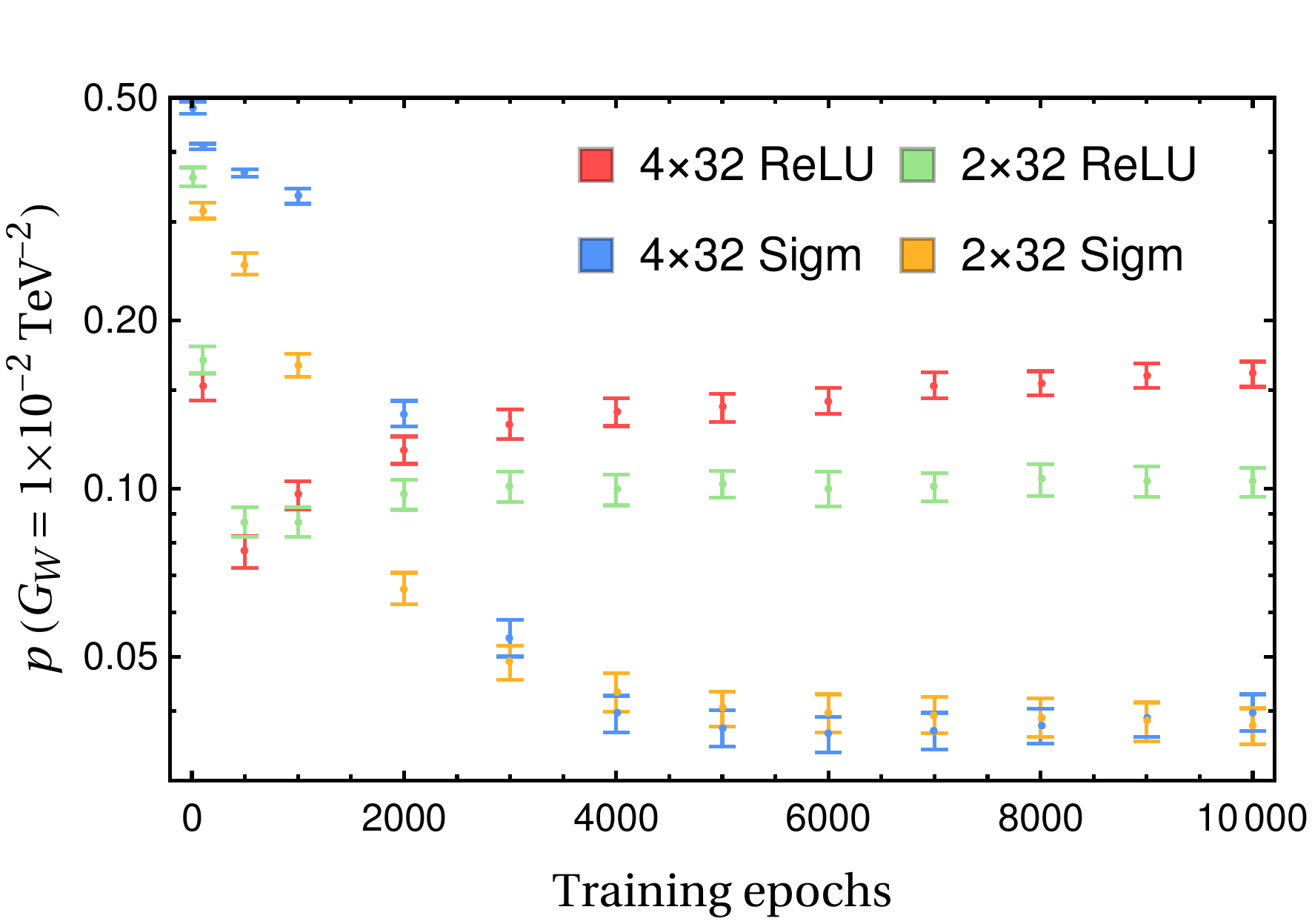}
	\caption{The $p$-value evolution during training for the Standard Classifier using different architectures and activation functions. The value $G_W=1\times10^{-2}\,{\rm TeV}^{-2}$ is employed.}
	\label{fig:nonparamevolution}
\end{figure}

\subsection{The Standard Classifier}

Hyperparameters optimization is rather different for the Standard Classifier. We see in Figure~\ref{fig:nonparamevolution} that Networks with ReLU activation like those we employed for the Quadratic Classifier displays overfitting, and Sigmoid activations need to be employed. The results in Figures~\ref{fig:fits_gphi} and in Table~\ref{tab:bounds} were obtained with $2$ hidden layers with $32$ neurons each and Sigmoid activation. The figure shows that increasing the complexity does not improve the performances. 

This different behavior of the Standard Classifier compared with the Quadratic one is probably due to the fact that training is performed on small Wilson coefficient EFT data, whose underlying distribution is very similar to the one of the SM data sample. Therefore there is not much genuine difference between the two training sets, and the Network is sensitive to statistical fluctuations in the training samples. The Quadratic Classifier instead is trained with large values of the Wilson coefficients. The optimizer thus drives the Neural Networks towards the deep minimum that corresponds to a proper modeling of the distribution ratio, which is more stable against statistical fluctuations of the training samples.

\subsection{Validation}

\begin{figure}[t]
	\centering
	\includegraphics[width=0.3\linewidth]{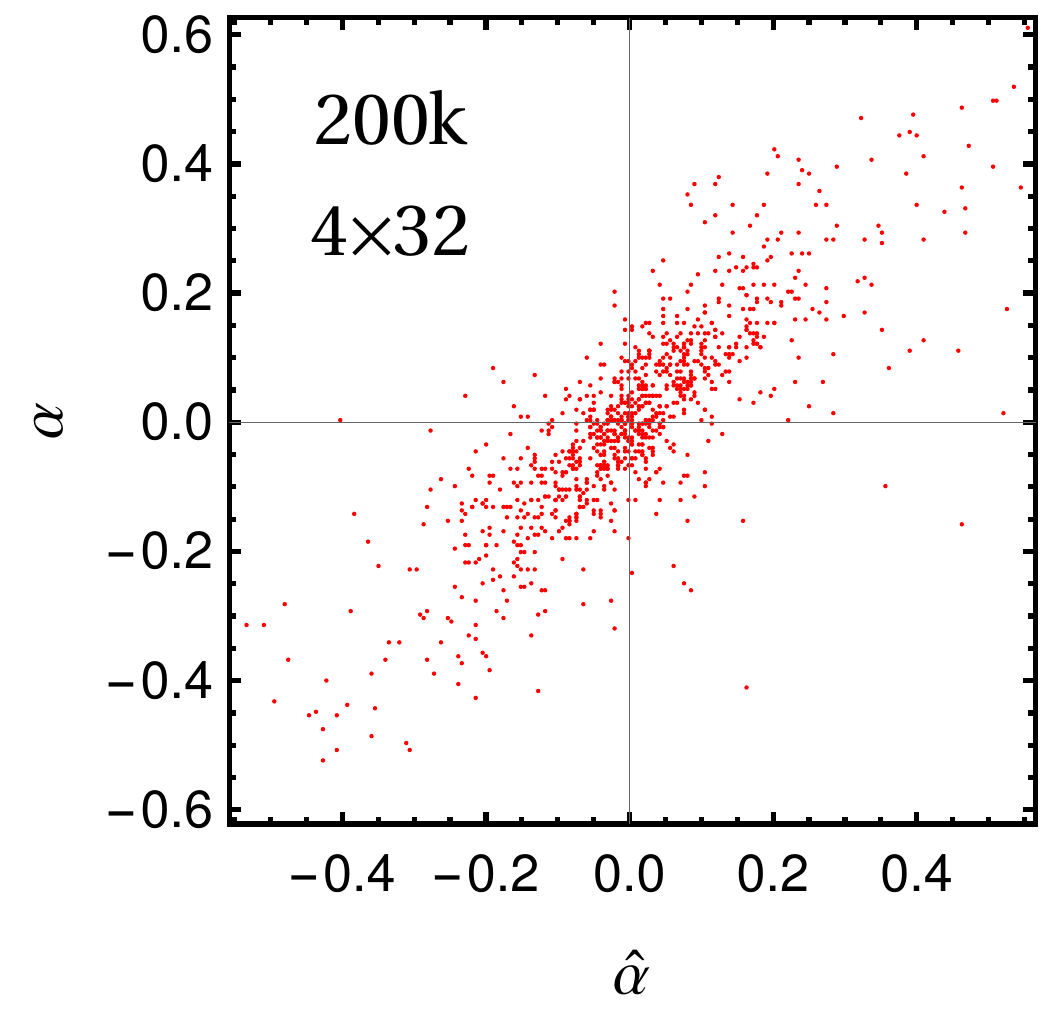}\hfill
	\includegraphics[width=0.3\linewidth]{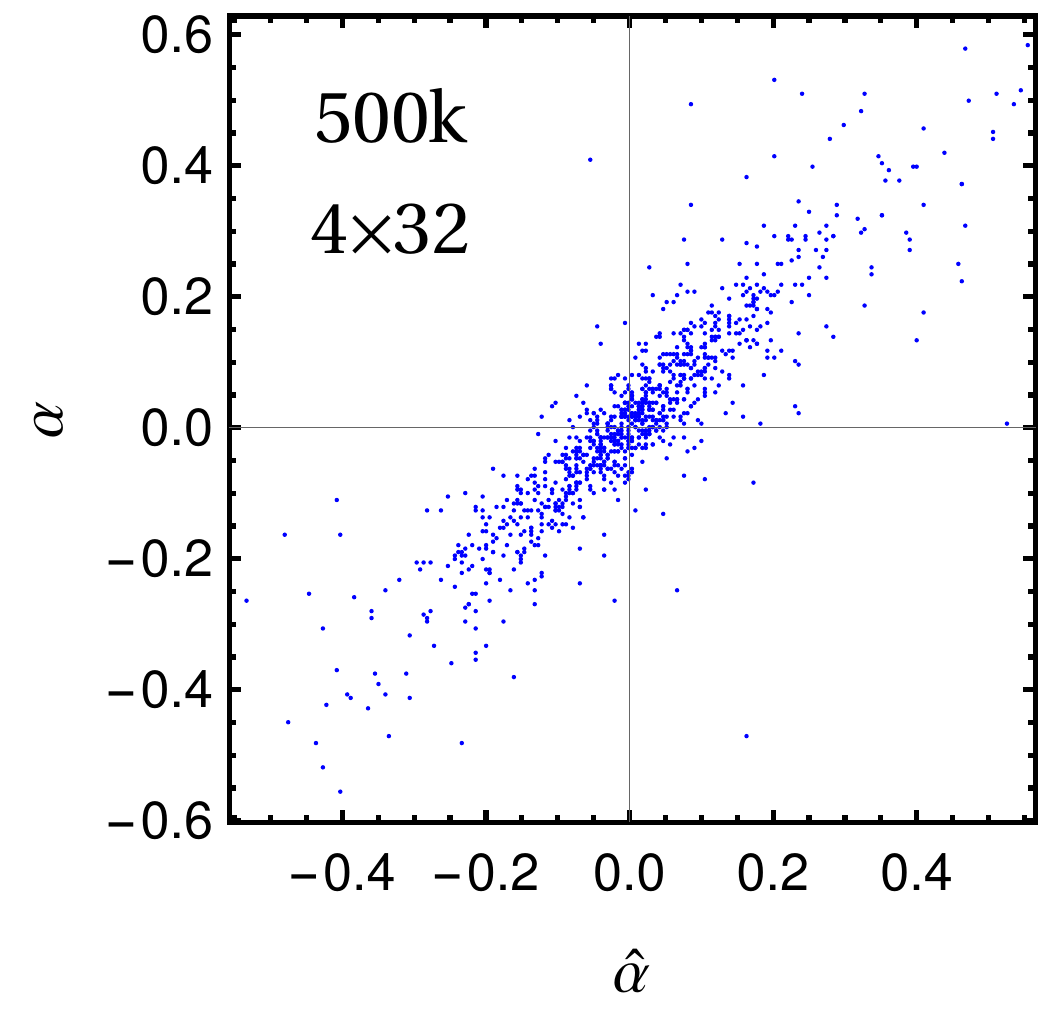}\hfill
	\includegraphics[width=0.3\linewidth]{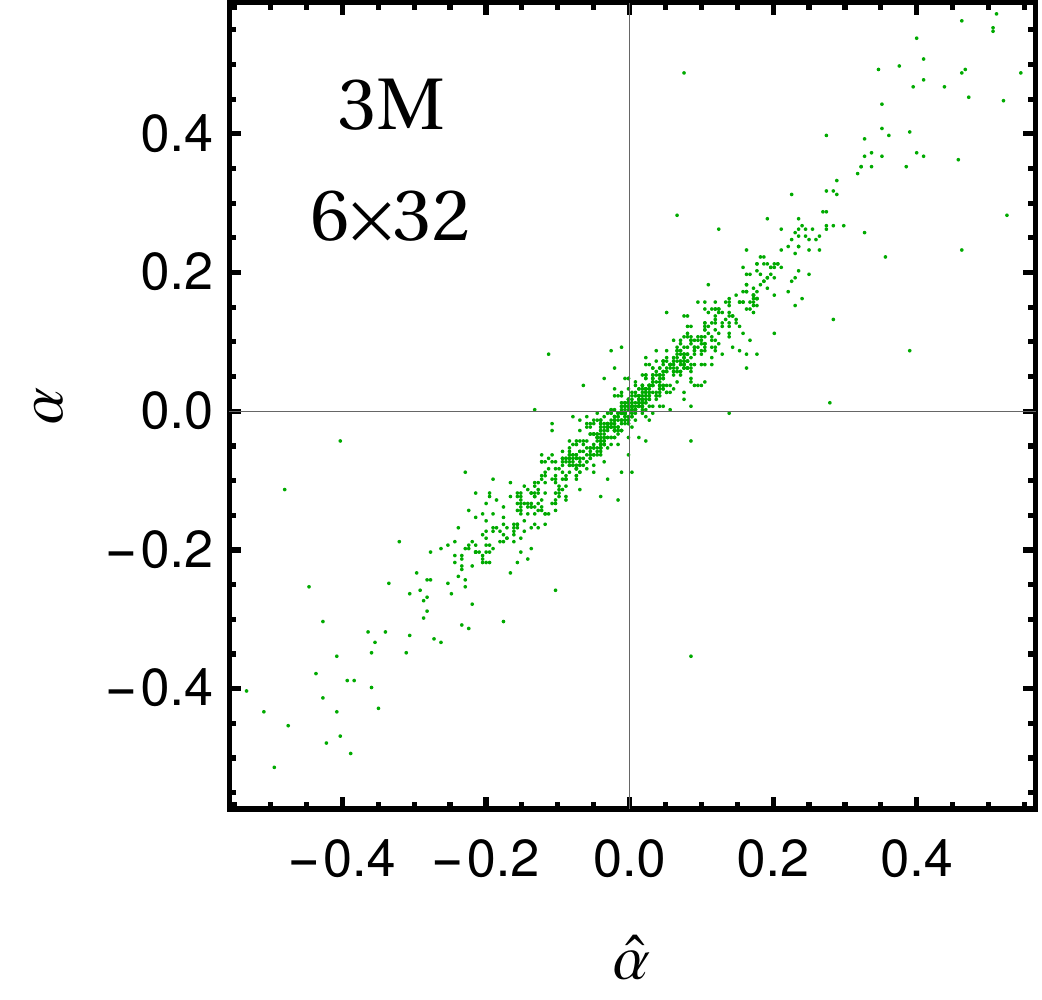}
	\caption{Comparison between the reconstructed ($\widehat{\alpha}$) and true ($\alpha$) linear term of the distribution ratio for the $G_W$ operator.}
	\label{fig:scatter}
\end{figure}

\begin{figure}
	\centering
	\includegraphics[width=0.3\linewidth]{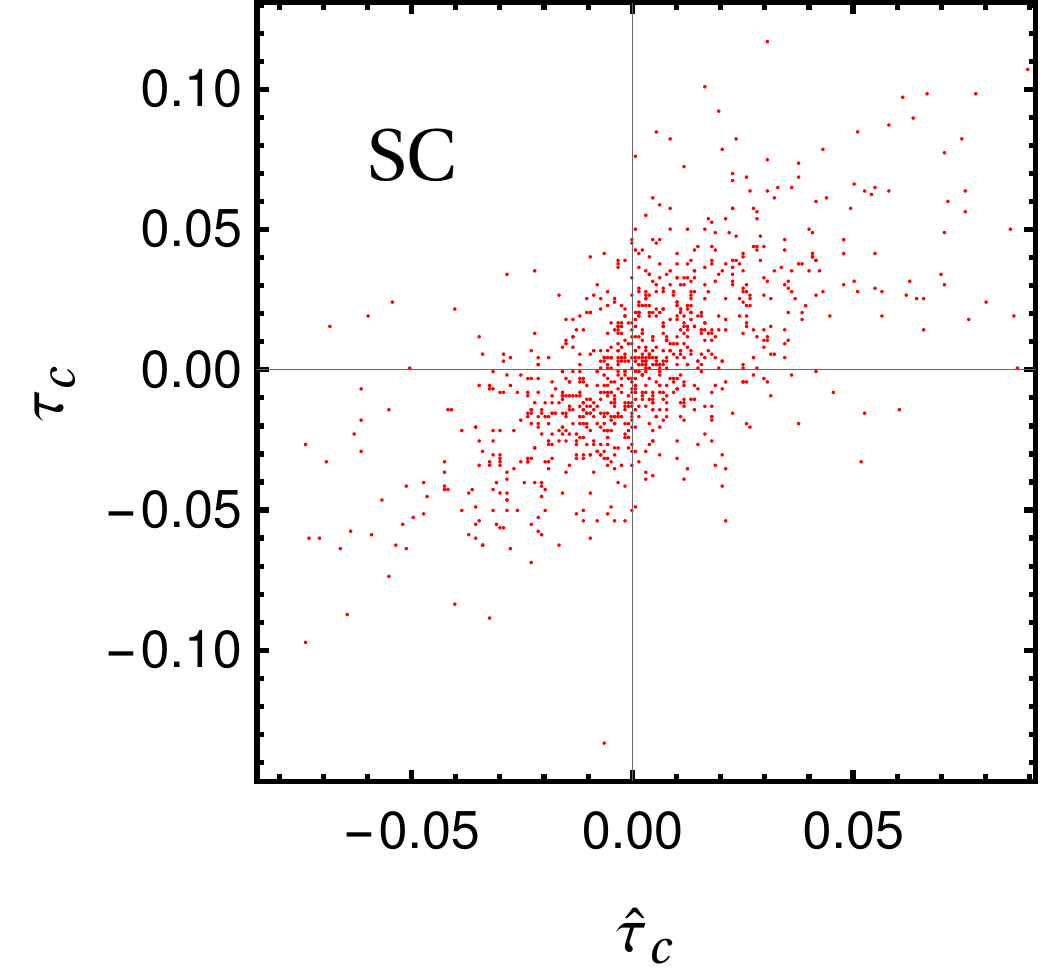}\hspace{20pt}
	\includegraphics[width=0.3\linewidth]{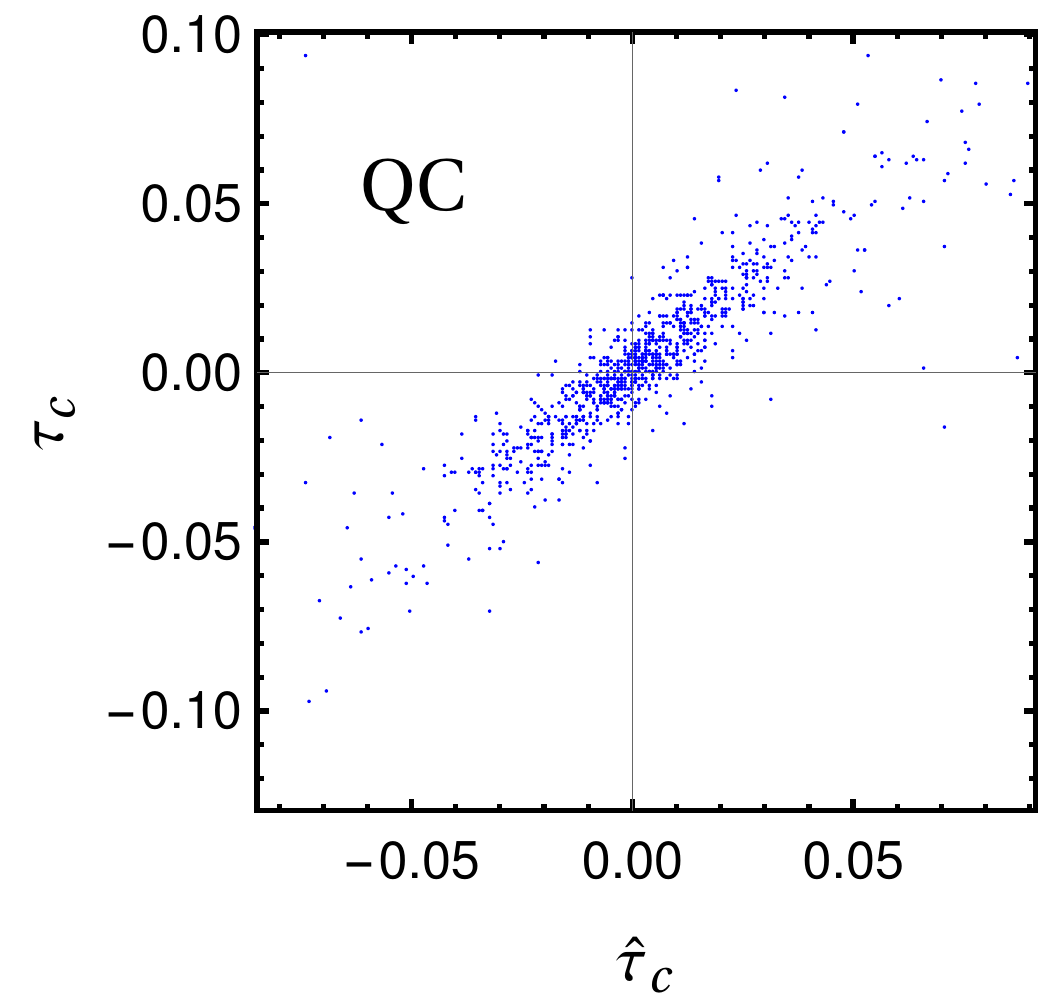}
	\caption{Comparison between the reconstructed ($\widehat{\tau}_c$) and true (${\tau}_c$) distribution log-ratio for $G_W=1\times10^{-2}\,{\rm TeV}^{-2}$. The Standard Classifier and the Quadratic one are considered in the left and right panel of the figure, respectively.}
	\label{fig:scatter2}
\end{figure}

An important question is how to validate as ``satisfactory'' the outcome of the hyperparameters optimization described above. This is straightforward for the Toy version of the problem, because we have to our disposal a rigorous notion of statistical optimality, through the Neyman--Pearson lemma, and we do have direct access to the true distribution ratio through which the data are generated. Therefore we know that we can stop optimization as soon as the reach of the Neural Network becomes sufficiently close to the one of the Matrix Element method. We can also rely on a more naive validation test, based on comparing point-by-point the distribution ratio learned by the Neural Network with the true one, which is known analytically. For instance in Figure~\ref{fig:scatter} we compare the true linear term $\alpha(x)$ in eq.~(\ref{QP}) (for the ${\mathcal{O}}_W$ operator) with its estimator $\widehat\alpha(x)\equiv\widehat{\rm{n}}_{\alpha}(x)$ provided by the trained Neural Network. The baseline architecture is employed, with increasing number of training points. While it is impossible to extract quantitative information, a qualitative comparison between the three scatter plots confirms that more training points improve the quality of the reconstruction. We also show, in Figure~\ref{fig:scatter2}, the correlation between the true and the reconstructed ratios (for $G_W=1\times10^{-2}\,{\rm TeV}^{-2}$, which corresponds to the Standard Classifier $95\%$ reach) obtained with the Quadratic and with the Standard Classifier. The reconstruction obtained with the Quadratic Classifier is more accurate as expected.

Validation is of course less easy if, as it is always the case on real problems, the true distribution ratio is not known. One option is to proceed like we did in the present paper. Namely to identify a Toy version of the problem that is sufficiently close to the real one and for which the distribution ratio is known. Since it is unlikely that the true distribution is much harder to learn than the Toy distribution, and since we can establish optimality on the Toy data using a certain architecture and training dataset size, we can argue heuristically that the same configuration will be optimal also with a more refined Monte Carlo description. 

\begin{figure}[t]
	\centering
	\includegraphics[width=0.3\linewidth]{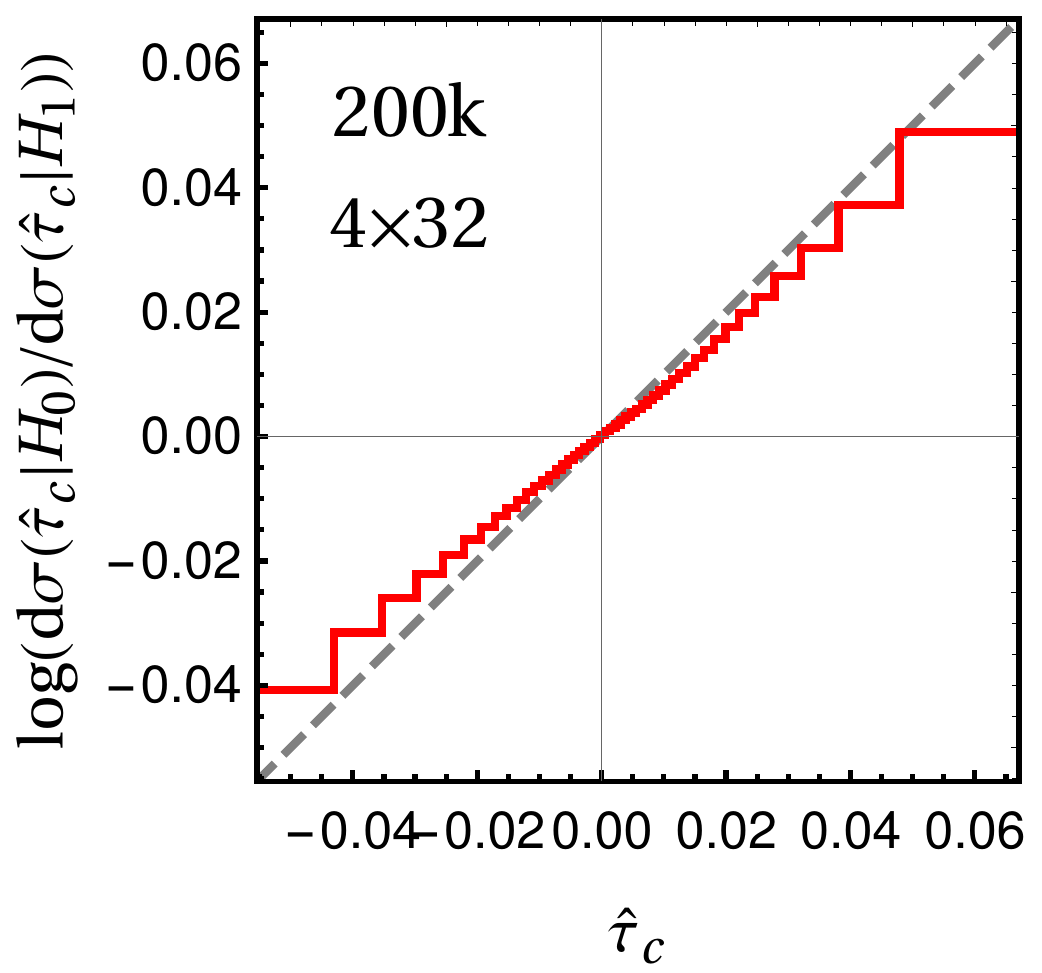}\hfill
	\includegraphics[width=0.3\linewidth]{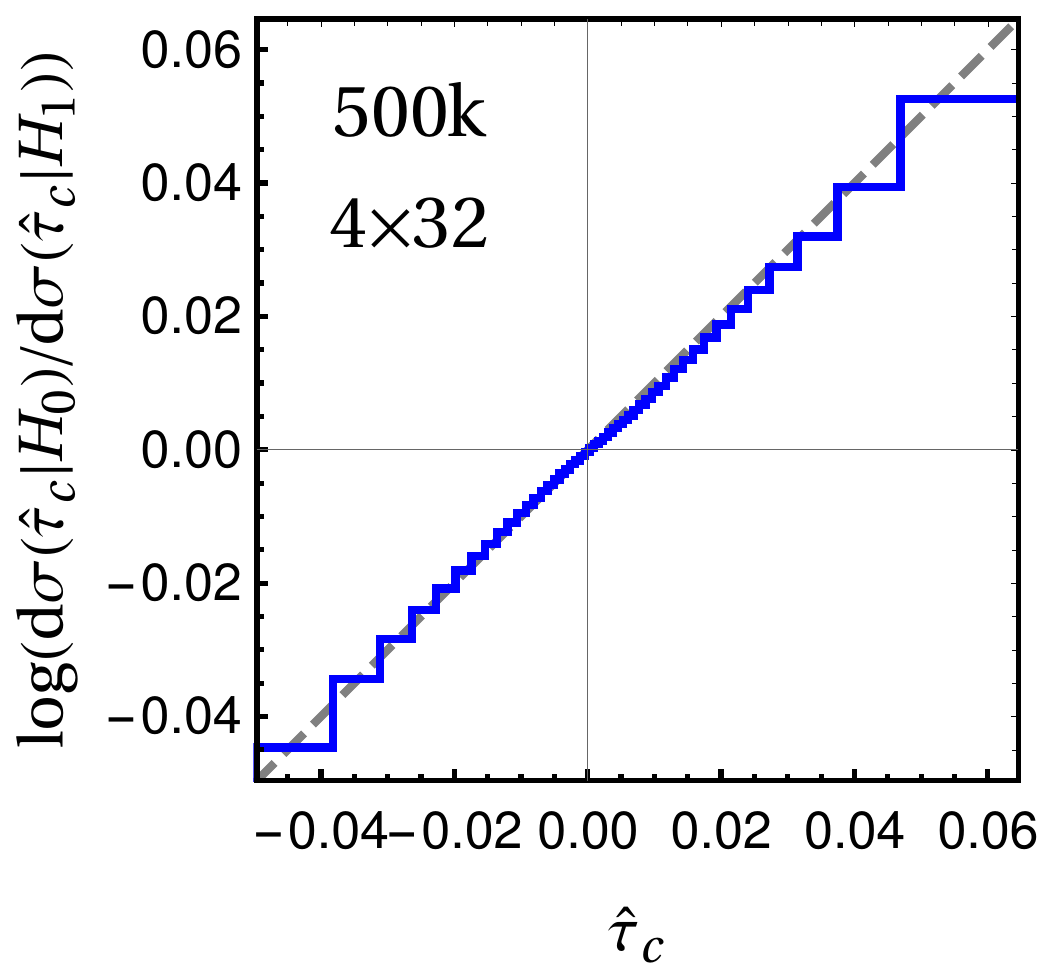}\hfill
	\includegraphics[width=0.3\linewidth]{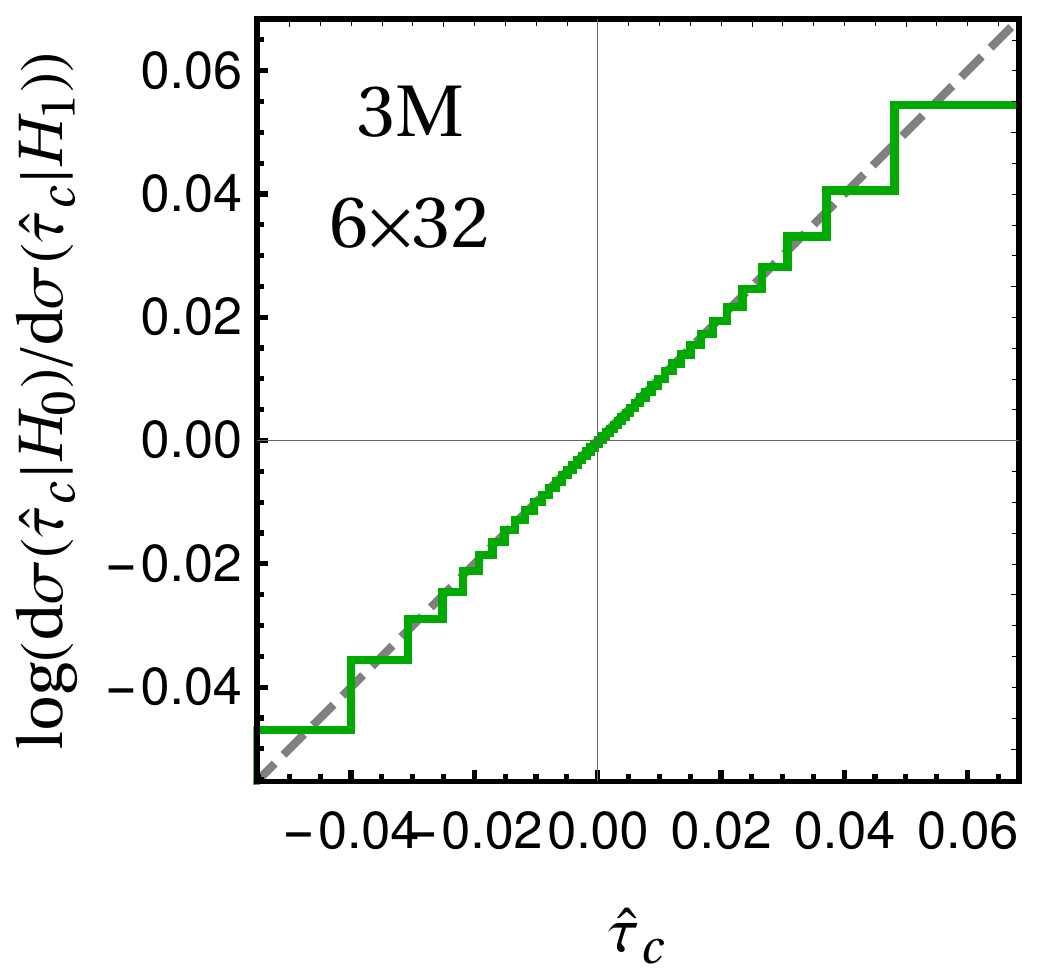}
	\caption{Distribution log-ratio for ${{\widehat{\tau}}_c}$, for $c=G_W=0.8\times10^{-2}\,{\rm TeV}^{-2}$. The accurate determination displayed in the plots is obtained by the reweighting of a single Toy SM Monte Carlo sample. The same approach, based on reweighting, could have been adopted to asses the quality of the distribution ratio reconstruction on {\sc MadGraph}~Monte Carlo data, using {\sc MadWeight}.}
	\label{fig:validation}
\end{figure}

Finally, one can monitor heuristically how accurately the distribution ratio is reconstructed, as follows. The true distribution log-ratio $\tau_c(x)=\log{r(x,c)}$, seen as a statistical variable  for each fixed value of $c$, obeys, by definition, the equation
\beq
\frac{d\sigma_0}{d\tau_c}=e^{\tau_c}\,\frac{d\sigma_1}{d\tau_c}\,.
\eeq
Therefore if we computed the distribution of $\tau_c$ (if it was known) in the EFT hypothesis $H_0(c)$ and in the SM hypothesis $H_1$, and take the log-ratio, the result would be a straight line as a function of $\tau_c$. By computing the same distributions for the reconstructed distribution log-ratio ${{\widehat{\tau}}_c}=\log{\widehat{r}}(x,c)$, we can thus get an indication of how closely ${\widehat{r}}(x,c)$ approximates $r(x,c)$. While no quantitative information can be extracted from these plots, they clearly illustrate the improvement achieved by enlarging the size of the training sample and the Neural Network architecture, as Figure~\ref{fig:validation} shows.

\section{Conclusions and outlook}\label{sec:conc}

We studied the potential gain in sensitivity of EFT searches at the LHC from multivariate analysis techniques. The results reported in Figure~\ref{fig:fits_gphiNLO} show that a considerable improvement is possible, especially for operators (like ${\mathcal{O}}_W$) with a complex interference pattern that is difficult to capture with a Binned Analysis.

Multivariate analyses based on Statistical Learning techniques are particularly promising, and should be considered as an alternative to the more standard (though not yet employed for EFT LHC searches) Matrix Element method. The advantage is eminently practical, because the Matrix Element method is optimal in principle, as much as the Statistical Learning approach. However the Matrix Element method needs to be designed case-by-case, and re-designed for each new effect one is willing to add for a more accurate modeling of the distribution ratio. It already required some effort to compute the approximate distribution in Section~\ref{sec:anap}, which in turn provides the simplest modeling of the distribution ratio to be employed in the Matrix Element approach, and we saw that this modeling is inadequate to describe the LO and even less adequate at NLO. In order to improve the modeling in the case at hand one should model the neutrino reconstruction more accurately, for instance by performing the integral over the neutrino momentum point-by-point in the space of the observed kinematical variables. The integral on the radiation should be also performed if willing to add QCD NLO effects. The predictions should be further refined including transfer functions for the detector effects, if the method has to be employed on real data.

The situation is radically different with the Statistical Learning approach. We saw that the exact same computational effort is required to reconstruct the distribution ratio at the Toy level, at LO and at NLO. Furthermore the accuracy of the reconstruction can be systematically improved using more training points and bigger Networks. The limiting factor is not reconstructing the distribution ratio by the Neural Network training. That step takes a quite small fraction of the computing time. The most time-consuming part of the procedure is the generation of the Monte Carlo training data, which becomes increasingly demanding as the sophistication of the Monte Carlo code increases. Even if we are still far from the limit for our analysis, it would be worth investigating improvements on this aspect based on Monte Carlo reweighting techniques.

It should be emphasized that Machine Learning methodologies are useful for EFT studies not only in view of the possible application to the analysis of the real data. After the conceptual and technical framework is in place, it is very easy to run the Machine Learning algorithm on the specific EFT problem at hand, and to get a feeling of the potential improvement of the reach compared with other methods. For instance our results show that the Binned Analysis we employed is inadequate for $G_W$, and that even for $G_{\varphi q}^{(3)}$ it could be improved. Furthermore they provide a target for the sensitivity such improvements should attain. Similarly, the results outline the importance of neutrino reconstruction modeling and of NLO QCD corrections being implemented in the Matrix Element method, if one is willing to adopt that strategy.

When it comes to the direct applicability of the method to the data, of the ZW process for instance, two additional steps are needed. The first one is to further improve the level of detail of the simulation. Detector effects could be added very easily with {\sc{Delphes}}~\cite{deFavereau:2013fsa}. However the reliability of the {\sc{Delphes}} description of the detectors should be cross-checked with a complete simulation by the experimental collaborations, and the {\sc{Delphes}} simulation replaced with a full detector simulation, which is much more demanding, if needed. 

The second aspect is to include systematic uncertainties of theoretical and experimental origin. It should be stressed that this is not more problematic in the Machine Learning framework than it is in the Matrix Element or any other multivariate approach. In particular it should be noticed that one has full control on the choice of the input variables that are given to the Neural Network and from which the sensitivity emerges. For instance in our case these would be the kinematical variables of the high-level reconstructed leptons, better if including photon recombination, in order to reduce the sensitivity to detector effects and showering, which might not be modeled accurately enough. Similarly if jets were used in the final state, high level IR-safe observables would be employed to be insensitive to hadronization, exactly like one would do for the Matrix Element method. It should also be stressed, as explained in the Introduction, that our method can be employed also in the presence of reducible backgrounds that must be extracted from the data because no reliable Monte Carlo generator is available.

The simplest strategy to deal with uncertainties is to merely quantify their impact on the sensitivity, using as discriminating variable the distribution ratio reconstructed from the nominal Monte Carlo generator that does not incorporate uncertainties. This is suboptimal, but sufficient to obtain conservative (i.e.~correct) results, and to identify the irrelevant sources of uncertainties. For better results one can include the uncertainties in the likelihood (i.e.~in the reconstructed distribution ratio) in the form of nuisance parameters. This is perfectly compatible with the Machine Learning approach, and already implemented in {\sc{MadMiner}}~\cite{Brehmer:2019xox} through morphing. Actually the Quadratic Classifier we employ in this paper could be useful also for this task. We will return to this point at the end of the Section. While conceptually straightforward, it is quantitatively important to assess the impact of uncertainties on the sensitivities we obtained in Figure~\ref{fig:fits_gphiNLO} on purely statistical grounds. This is left to future work.

One interesting technical element of the present paper is the Quadratic Classifier, introduced in Section~\ref{PaC}. We have found that it performs better than the Standard Classifier, as expected since it is designed to be sensitive to the small departures from the SM due to the EFT by exploiting the exact knowledge of the (quadratic) functional dependence of the distribution ratio on the Wilson coefficients. Furthermore it is computationally much more convenient and thus feasible also when several EFT operators are considered simultaneously and the scan over the Wilson coefficients becomes unfeasible. The Quadratic Classifier has been found to be nearly optimal, with a rigorous notion of optimality based on the Neyman--Pearson lemma. 

We described in the body of the paper the connection between the Quadratic Classifier and other techniques based on Statistical Learning available in the literature, but we did not yet discuss the relation with the most sophisticated such techniques, namely the ones that exploit ``hidden'' information from the Monte Carlo simulator~\cite{Brehmer:2018hga}. The basic idea is that the simulator does contain the analytic information on the underlying distribution, and so it does contain a representation of the EFT/SM distribution ratio in terms of latent variables. One can incorporate this information in the loss function, so that the machine does not need to learn the likelihood ratio from scratch, but only the distortions of the likelihood ratio due to the transition between the latent and the true variables. The Quadratic Classifier trick is orthogonal to this interesting idea, and it could be straightforwardly implemented in the simulator-assisted methods by modifying the loss function in close analogy with eq.~(\ref{LFPC}). The advantages of parametrization in that context could be the same we observed here. 

On the other hand, simulator-assisted methods have also potential limitations, in two respects. First, because there is a clear benefit from exploiting the latent-space distribution ratio if the latter is similar to the one in the space of observables, but this is not necessarily the case. For instance in ZW we saw that a proper modeling of the neutrino reconstruction is crucial for the performances, and this is not captured by the latent-variables ratio that involves the true neutrino momentum. This can be a problem for the validation of the approach, due to the fact that any additional effect we include in the simulation, which further distorts the observed ratio, might be more and more difficult for the machine to learn. For instance a simulator-assisted method should be trivially optimal on the Toy data, where the latent space coincides with the observed space and thus the likelihood ratio employed in training coincides with the true one and the machine has nothing to learn. However this does not mean that it will work on the LO data (using the appropriate LO latent-variables ratio) because now the machine has the non-trivial task to integrate out the neutrino. Instead for our method, that learns the distribution ratio using no information from the Monte Carlo apart from the event sample itself, it is arguably equally difficult to model the distribution ratio on the Toy, on the LO and on the NLO data. Therefore the optimality on Toy data, which we can establish rigorously because we know the exact distribution ratio, heuristically indicates that the algorithm is optimal at LO and NLO as well. The second problem of simulator-assisted method is that the required information on the latent-space distribution ratio might not be made available by the Monte Carlo code. In light of this, it is reassuring to have an alternative method that does not rely on latent-space information, that is feasible and optimal, at least in the case at hand.

Finally, it should be noticed that the parametrization trick is not specific of the EFT and it could be applied to any situation where the functional dependence of the distribution on the parameters is either exactly or approximately known. One should just replace the quadratic dependence of eq.~(\ref{PC}) on $c$ with the appropriate (polynomial or not) functional form. This could be useful to include the effect of nuisance parameters in the likelihood. Nuisance parameters effects on the distribution can be normally modeled linearly (or with an exponential, to avoid negative distributions) to good approximation because their effects are small. However if they are too small (but still potentially competitive with the EFT ones) it could be difficult for the machine to learn them using simulations where the nuisances are varied within their one-sigma interval. If the analytic dependence on the nuisance parameters is incorporated  in the classifier, we could ameliorate the situation by training with larger values of the parameters like we did in this paper to reconstruct the small EFT effects. Exploring this direction is left to future work. 

\section*{Acknowledgments}
We thank J.~Brehmer and K.~Cranmer for useful discussions. The work of S.C. was supported by the Swiss National Science Foundation under contract 200021-178999. The Swiss National Science Foundation supported the work of A.G. under contracts 200020-169696. G.P.~was supported in part by the MIUR under contract 2017FMJFMW (PRIN2017).

\appendix

\section{The general Quadratic Classifier}\label{app:A}

Any quadratic-order real polynomial of $n-1$ variables $c_i$, $i=1,\ldots n-1$, with arbitrary constant, linear and quadratic terms, can be written as a quadratic form in the $n$-dimensional variable 
\beq
v(c)=(1,\,c_1,\,\ldots,\,c_{n-1})^T\,.
\eeq
Namely, we write the polynomial as
\beq\label{polgen}
P(c)=v^T(c)A\,v(c)\,,
\eeq
with $A$ a generic $n$-dimensional real symmetric square matrix. 

If $P(c)$ is non-negative for any value of $c$, it is easy to show that the matrix $A$ must be positive semi-definite. 
Being real, symmetric and positive semi-definite, it is possible to use the Cholesky decomposition for $A$, and write it as
\beq
A=L^T L\,,
\eeq
where $L$ is a upper-triangular (i.e., $L_{ij}=0$ for $j<i$) real matrix. Therefore the most general positive quadratic order polynomial reads
\beq\label{CD}
P(c)=v^T(c)L^T L\,v(c)=\sum\limits_{i=1}^n\left(\sum\limits_{j=1}^n L_{ij}v_j(c)\right)^2=\sum\limits_{i=1}^n\left(L_{i1}+\sum\limits_{j=2}^{n} L_{ij}c_{j-1}\right)^2\,,
\eeq
which is manifestly non-negative because it is the sum of square terms. Moreover for $c=0$, since $L_{i1}=L_{11}\delta_{i1}$, we have $P(0)=L_{11}^2$. The Cholesky decomposition is unique up to sign flips of the rows of $L$. Rather than resolving this ambiguity, for instance by choosing the diagonal entries of $L$ to be positive, we adopt eq.~(\ref{CD}) without further constraints as the most general (though redundant) parametrization of $P(c)$.

The EFT differential cross section is a positive quadratic polynomial in the Wilson Coefficient $c_i$ at each phase-space point $x$, and it reduces to the SM cross section for $c=0$. It must therefore take the form
\beq\label{ratioGEN}
d\sigma_0(x;c)=d\sigma_1(x)\sum\limits_{i=1}^n\left[\delta_{i1}+\sum\limits_{j=2}^{n} \lambda(x)_{ij}c_{j-1}\right]^2\,,
\eeq
with $\lambda(x)$ an upper-triangular matrix of real functions.
If only one $c$ parameter is present (i.e., $n=2$), this reduces to eq.~(\ref{QP}) with the identifications
\beq
\lambda(x)_{12}=\alpha(x) \;\;\;\;\;\lambda(x)_{22}=\beta(x)\,.
\eeq
The Quadratic Classifier that generalizes eq.~(\ref{PC}) is thus defined as
\beq\label{GPC}
f(x,c)\equiv\frac1{1+\sum\limits_{i=1}^n\left[\delta_{i1}+\sum\limits_{j=2}^{n} {\rm{n}}(x)_{ij}c_{j-1}\right]^2}\,,
\eeq
in terms of an upper-triangular matrix ${\rm{n}}(x)$ of real-output Neural Networks.

\section{Minimization of the parametrized loss}\label{app:B}

In the Large Sample limit, the loss function in eq.~(\ref{LFPC}) becomes
\beq
L[{\rm{n}}(\cdot)]\overset{\rm\sc{LS}}{=}\sum\limits_{c\in{\mathcal{C}}}\left\{ \int \hspace{-3pt} d\sigma_0(x;c) [f(x,c)]^2 +\int\hspace{-3pt} d\sigma_1(x)[1-f(x,c)]^2\right\}\,,
\eeq
with the Quadratic Classifier $f$ defined in eq.~(\ref{GPC}). By simple algebraic manipulations, this can be rewritten as
\beq
L[{\rm{n}}(\cdot)]\overset{\rm\sc{LS}}{=}\sum\limits_{c\in{\mathcal{C}}}\left\{ 
\int \hspace{-3pt} \frac{d\sigma_1(x)d\sigma_0(x;c)}{d\sigma_1(x)+d\sigma_0(x;c)}
 +\int\hspace{-3pt} [d\sigma_1(x)+d\sigma_0(x;c)] \left[f(x,c)-\frac1{1+r(x,c)} \right]^2\right\}\,,
\eeq
with $r(x,c)=d\sigma_0(x;c)/d\sigma_1(x)$. The first integral is independent of $f$ and thus it is irrelevant for the minimization of the loss. The second one is the integral of a non-negative function of $x$ which attains its global minimum (i.e., it vanishes) if and only if
\beq
f(x,c)=f_{\rm{min}}(x,c)=\frac1{1+r(x,c)}\,,\;\;\;\;\;\forall \,c\in{\mathcal{C}}\,.
\eeq
By using eq.~(\ref{ratioGEN}), and comparing with eq.~(\ref{GPC}), we immediately conclude that the configuration ${\rm{n}}(x)_{ij}=\lambda(x)_{ij}$ is a global minimum of the loss and that this minimum is unique provided the set ${\mathcal{C}}$ contains at least two distinct non-vanishing elements. More precisely, this holds only up to sign ambiguities, associated with  those of the Cholesky decomposition. However this is irrelevant because the ambiguity cancels out in $f$, and in turn it cancels out in the reconstructed distribution ratio $\widehat{r}(x,c)=1/\widehat{f}(x,c)-1$.

We have shown that the Quadratic Classifier reconstructs the distribution ratio exactly (in the Large Sample limit and for infinitely complex Neural Network) at the global minimum of the loss, and that this minimum is unique. Notice however that we could not show that the Large Sample limit loss does not possess additional local minimums, as it is instead readily proven for the standard classifier of Section~\ref{SC} by variational calculus.

\bibliographystyle{JHEP}

\begin{thebibliography}{10}

\bibitem{Buchmuller:1985jz}
W.~Buchmuller and D.~Wyler, \emph{{Effective Lagrangian Analysis of New
  Interactions and Flavor Conservation}},
  \href{https://doi.org/10.1016/0550-3213(86)90262-2}{\emph{Nucl. Phys. B}
  {\bfseries 268} (1986) 621}.

\bibitem{Giudice:2007fh}
G.~Giudice, C.~Grojean, A.~Pomarol and R.~Rattazzi, \emph{{The
  Strongly-Interacting Light Higgs}},
  \href{https://doi.org/10.1088/1126-6708/2007/06/045}{\emph{JHEP} {\bfseries
  06} (2007) 045} [\href{https://arxiv.org/abs/hep-ph/0703164}{{\ttfamily
  hep-ph/0703164}}].

\bibitem{Grzadkowski:2010es}
B.~Grzadkowski, M.~Iskrzynski, M.~Misiak and J.~Rosiek, \emph{{Dimension-Six
  Terms in the Standard Model Lagrangian}},
  \href{https://doi.org/10.1007/JHEP10(2010)085}{\emph{JHEP} {\bfseries 10}
  (2010) 085} [\href{https://arxiv.org/abs/1008.4884}{{\ttfamily 1008.4884}}].

\bibitem{Atwood:1991ka}
D.~Atwood and A.~Soni, \emph{{Analysis for magnetic moment and electric dipole
  moment form-factors of the top quark via $e^+e^-\rightarrow
  t{\overline{t}}$}},
  \href{https://doi.org/10.1103/PhysRevD.45.2405}{\emph{Phys. Rev. D}
  {\bfseries 45} (1992) 2405}.

\bibitem{Diehl:1993br}
M.~Diehl and O.~Nachtmann, \emph{{Optimal observables for the measurement of
  three gauge boson couplings in $e^+ e^- \rightarrow W^+ W^-$}},
  \href{https://doi.org/10.1007/BF01555899}{\emph{Z. Phys. C} {\bfseries 62}
  (1994) 397}.

\bibitem{Dunietz:1990cj}
I.~Dunietz, H.~R. Quinn, A.~Snyder, W.~Toki and H.~J. Lipkin, \emph{{How to
  extract CP violating asymmetries from angular correlations}},
  \href{https://doi.org/10.1103/PhysRevD.43.2193}{\emph{Phys. Rev. D}
  {\bfseries 43} (1991) 2193}.

\bibitem{Dighe:1998vk}
A.~S. Dighe, I.~Dunietz and R.~Fleischer, \emph{{Extracting CKM phases and $B_s
  - \bar{B}_s$ mixing parameters from angular distributions of nonleptonic $B$
  decays}}, \href{https://doi.org/10.1007/s100520050372}{\emph{Eur. Phys. J. C}
  {\bfseries 6} (1999) 647}
  [\href{https://arxiv.org/abs/hep-ph/9804253}{{\ttfamily hep-ph/9804253}}].

\bibitem{Banerjee:2019twi}
S.~Banerjee, R.~S. Gupta, J.~Y. Reiness, S.~Seth and M.~Spannowsky,
  \emph{{Towards the ultimate differential SMEFT analysis}},
  \href{https://arxiv.org/abs/1912.07628}{{\ttfamily 1912.07628}}.

\bibitem{Anderson:2013afp}
I.~Anderson et~al., \emph{{Constraining Anomalous HVV Interactions at Proton
  and Lepton Colliders}},
  \href{https://doi.org/10.1103/PhysRevD.89.035007}{\emph{Phys. Rev. D}
  {\bfseries 89} (2014) 035007}
  [\href{https://arxiv.org/abs/1309.4819}{{\ttfamily 1309.4819}}].

\bibitem{Kondo:1988yd}
K.~Kondo, \emph{{Dynamical Likelihood Method for Reconstruction of Events With
  Missing Momentum. 1: Method and Toy Models}},
  \href{https://doi.org/10.1143/JPSJ.57.4126}{\emph{J. Phys. Soc. Jap.}
  {\bfseries 57} (1988) 4126}.

\bibitem{Artoisenet:2010cn}
P.~Artoisenet, V.~Lemaitre, F.~Maltoni and O.~Mattelaer, \emph{{Automation of
  the matrix element reweighting method}},
  \href{https://doi.org/10.1007/JHEP12(2010)068}{\emph{JHEP} {\bfseries 12}
  (2010) 068} [\href{https://arxiv.org/abs/1007.3300}{{\ttfamily 1007.3300}}].

\bibitem{Fiedler:2010sg}
F.~Fiedler, A.~Grohsjean, P.~Haefner and P.~Schieferdecker, \emph{{The Matrix
  Element Method and its Application in Measurements of the Top Quark Mass}},
  \href{https://doi.org/10.1016/j.nima.2010.09.024}{\emph{Nucl. Instrum. Meth.
  A} {\bfseries 624} (2010) 203}
  [\href{https://arxiv.org/abs/1003.1316}{{\ttfamily 1003.1316}}].

\bibitem{Martini:2015fsa}
T.~Martini and P.~Uwer, \emph{{Extending the Matrix Element Method beyond the
  Born approximation: Calculating event weights at next-to-leading order
  accuracy}}, \href{https://doi.org/10.1007/JHEP09(2015)083}{\emph{JHEP}
  {\bfseries 09} (2015) 083}
  [\href{https://arxiv.org/abs/1506.08798}{{\ttfamily 1506.08798}}].

\bibitem{Martini:2017ydu}
T.~Martini and P.~Uwer, \emph{{The Matrix Element Method at next-to-leading
  order QCD for hadronic collisions: Single top-quark production at the LHC as
  an example application}},
  \href{https://doi.org/10.1007/JHEP05(2018)141}{\emph{JHEP} {\bfseries 05}
  (2018) 141} [\href{https://arxiv.org/abs/1712.04527}{{\ttfamily
  1712.04527}}].

\bibitem{Brehmer:2018kdj}
J.~Brehmer, K.~Cranmer, G.~Louppe and J.~Pavez, \emph{{Constraining Effective
  Field Theories with Machine Learning}},
  \href{https://doi.org/10.1103/PhysRevLett.121.111801}{\emph{Phys. Rev. Lett.}
  {\bfseries 121} (2018) 111801}
  [\href{https://arxiv.org/abs/1805.00013}{{\ttfamily 1805.00013}}].

\bibitem{Cranmer:2015bka}
K.~Cranmer, J.~Pavez and G.~Louppe, \emph{{Approximating Likelihood Ratios with
  Calibrated Discriminative Classifiers}},
  \href{https://arxiv.org/abs/1506.02169}{{\ttfamily 1506.02169}}.

\bibitem{Baldi:2016fzo}
P.~Baldi, K.~Cranmer, T.~Faucett, P.~Sadowski and D.~Whiteson,
  \emph{{Parameterized neural networks for high-energy physics}},
  \href{https://doi.org/10.1140/epjc/s10052-016-4099-4}{\emph{Eur. Phys. J. C}
  {\bfseries 76} (2016) 235}
  [\href{https://arxiv.org/abs/1601.07913}{{\ttfamily 1601.07913}}].

\bibitem{Stoye:2018ovl}
M.~Stoye, J.~Brehmer, G.~Louppe, J.~Pavez and K.~Cranmer,
  \emph{{Likelihood-free inference with an improved cross-entropy estimator}},
  \href{https://arxiv.org/abs/1808.00973}{{\ttfamily 1808.00973}}.

\bibitem{Brehmer:2018hga}
J.~Brehmer, G.~Louppe, J.~Pavez and K.~Cranmer, \emph{{Mining gold from
  implicit models to improve likelihood-free inference}},
  \href{https://doi.org/10.1073/pnas.1915980117}{\emph{Proc. Nat. Acad. Sci.}
  (2020) 201915980} [\href{https://arxiv.org/abs/1805.12244}{{\ttfamily
  1805.12244}}].

\bibitem{Brehmer:2018eca}
J.~Brehmer, K.~Cranmer, G.~Louppe and J.~Pavez, \emph{{A Guide to Constraining
  Effective Field Theories with Machine Learning}},
  \href{https://doi.org/10.1103/PhysRevD.98.052004}{\emph{Phys. Rev. D}
  {\bfseries 98} (2018) 052004}
  [\href{https://arxiv.org/abs/1805.00020}{{\ttfamily 1805.00020}}].

\bibitem{Brehmer:2019xox}
J.~Brehmer, F.~Kling, I.~Espejo and K.~Cranmer, \emph{{MadMiner: Machine
  learning-based inference for particle physics}},
  \href{https://doi.org/10.1007/s41781-020-0035-2}{\emph{Comput. Softw. Big
  Sci.} {\bfseries 4} (2020) 3}
  [\href{https://arxiv.org/abs/1907.10621}{{\ttfamily 1907.10621}}].

\bibitem{Neyman:1933wgr}
J.~Neyman and E.~S. Pearson, \emph{{On the Problem of the Most Efficient Tests
  of Statistical Hypotheses}},
  \href{https://doi.org/10.1098/rsta.1933.0009}{\emph{Phil. Trans. Roy. Soc.
  Lond. A} {\bfseries 231} (1933) 289}.

\bibitem{Tanabashi:2018oca}
{\scshape Particle Data Group} collaboration, M.~Tanabashi et~al.,
  \emph{{Review of Particle Physics}},
  \href{https://doi.org/10.1103/PhysRevD.98.030001}{\emph{Phys. Rev. D}
  {\bfseries 98} (2018) 030001}.

\bibitem{Falkowski:2015jaa}
A.~Falkowski, M.~Gonzalez-Alonso, A.~Greljo and D.~Marzocca, \emph{{Global
  constraints on anomalous triple gauge couplings in effective field theory
  approach}}, \href{https://doi.org/10.1103/PhysRevLett.116.011801}{\emph{Phys.
  Rev. Lett.} {\bfseries 116} (2016) 011801}
  [\href{https://arxiv.org/abs/1508.00581}{{\ttfamily 1508.00581}}].

\bibitem{Green:2016trm}
D.~R. Green, P.~Meade and M.-A. Pleier, \emph{{Multiboson interactions at the
  LHC}}, \href{https://doi.org/10.1103/RevModPhys.89.035008}{\emph{Rev. Mod.
  Phys.} {\bfseries 89} (2017) 035008}
  [\href{https://arxiv.org/abs/1610.07572}{{\ttfamily 1610.07572}}].

\bibitem{Butter:2016cvz}
A.~Butter, O.~J.~P. \'{E}boli, J.~Gonzalez-Fraile, M.~Gonzalez-Garcia, T.~Plehn
  and M.~Rauch, \emph{{The Gauge-Higgs Legacy of the LHC Run I}},
  \href{https://doi.org/10.1007/JHEP07(2016)152}{\emph{JHEP} {\bfseries 07}
  (2016) 152} [\href{https://arxiv.org/abs/1604.03105}{{\ttfamily
  1604.03105}}].

\bibitem{Franceschini:2017xkh}
R.~Franceschini, G.~Panico, A.~Pomarol, F.~Riva and A.~Wulzer,
  \emph{{Electroweak Precision Tests in High-Energy Diboson Processes}},
  \href{https://doi.org/10.1007/JHEP02(2018)111}{\emph{JHEP} {\bfseries 02}
  (2018) 111} [\href{https://arxiv.org/abs/1712.01310}{{\ttfamily
  1712.01310}}].

\bibitem{Panico:2017frx}
G.~Panico, F.~Riva and A.~Wulzer, \emph{{Diboson Interference Resurrection}},
  \href{https://doi.org/10.1016/j.physletb.2017.11.068}{\emph{Phys. Lett.}
  {\bfseries B776} (2018) 473}
  [\href{https://arxiv.org/abs/1708.07823}{{\ttfamily 1708.07823}}].

\bibitem{Azatov:2017kzw}
A.~Azatov, J.~Elias-Miro, Y.~Reyimuaji and E.~Venturini, \emph{{Novel
  measurements of anomalous triple gauge couplings for the LHC}},
  \href{https://doi.org/10.1007/JHEP10(2017)027}{\emph{JHEP} {\bfseries 10}
  (2017) 027} [\href{https://arxiv.org/abs/1707.08060}{{\ttfamily
  1707.08060}}].

\bibitem{Azatov:2019xxn}
A.~Azatov, D.~Barducci and E.~Venturini, \emph{{Precision diboson measurements
  at hadron colliders}},
  \href{https://doi.org/10.1007/JHEP04(2019)075}{\emph{JHEP} {\bfseries 04}
  (2019) 075} [\href{https://arxiv.org/abs/1901.04821}{{\ttfamily
  1901.04821}}].

\bibitem{Baglio:2019uty}
J.~Baglio, S.~Dawson and S.~Homiller, \emph{{QCD corrections in Standard Model
  EFT fits to $WZ$ and $WW$ production}},
  \href{https://doi.org/10.1103/PhysRevD.100.113010}{\emph{Phys. Rev. D}
  {\bfseries 100} (2019) 113010}
  [\href{https://arxiv.org/abs/1909.11576}{{\ttfamily 1909.11576}}].

\bibitem{Duncan:1985ij}
M.~J. Duncan, G.~L. Kane and W.~W. Repko, \emph{{A New Standard Model Test for
  Future Colliders}},
  \href{https://doi.org/10.1103/PhysRevLett.55.773}{\emph{Phys. Rev. Lett.}
  {\bfseries 55} (1985) 773}.

\bibitem{Hagiwara:1986vm}
K.~Hagiwara, R.~Peccei, D.~Zeppenfeld and K.~Hikasa, \emph{{Probing the Weak
  Boson Sector in $e^+ e^- \rightarrow W^+ W^-$}},
  \href{https://doi.org/10.1016/0550-3213(87)90685-7}{\emph{Nucl. Phys. B}
  {\bfseries 282} (1987) 253}.

\bibitem{Cowan:1998ji}
G.~Cowan, \emph{{Statistical data analysis}}. Oxford University Press, USA,
  1998.

\bibitem{NNL}
M.~Anthony and P.~L. Bartlett, \emph{{Neural Network Learning: Theoretical
  Foundations}}. Cambridge University Press., 1999.

\bibitem{DAgnolo:2018cun}
R.~T. D'Agnolo and A.~Wulzer, \emph{{Learning New Physics from a Machine}},
  \href{https://doi.org/10.1103/PhysRevD.99.015014}{\emph{Phys. Rev. D}
  {\bfseries 99} (2019) 015014}
  [\href{https://arxiv.org/abs/1806.02350}{{\ttfamily 1806.02350}}].

\bibitem{Cuomo:2019siu}
G.~Cuomo, L.~Vecchi and A.~Wulzer, \emph{{Goldstone Equivalence and High Energy
  Electroweak Physics}},
  \href{https://doi.org/10.21468/SciPostPhys.8.5.078}{\emph{SciPost Phys.}
  {\bfseries 8} (2020) 078} [\href{https://arxiv.org/abs/1911.12366}{{\ttfamily
  1911.12366}}].

\bibitem{Kusina:2015vfa}
A.~Kusina et~al., \emph{{nCTEQ15 - Global analysis of nuclear parton
  distributions with uncertainties}},
  \href{https://doi.org/10.22323/1.247.0041}{\emph{PoS} {\bfseries DIS2015}
  (2015) 041} [\href{https://arxiv.org/abs/1509.01801}{{\ttfamily
  1509.01801}}].

\bibitem{Clark:2016jgm}
D.~Clark, E.~Godat and F.~Olness, \emph{{ManeParse : A Mathematica reader for
  Parton Distribution Functions}},
  \href{https://doi.org/10.1016/j.cpc.2017.03.004}{\emph{Comput. Phys. Commun.}
  {\bfseries 216} (2017) 126}
  [\href{https://arxiv.org/abs/1605.08012}{{\ttfamily 1605.08012}}].

\bibitem{Alwall:2014hca}
J.~Alwall, R.~Frederix, S.~Frixione, V.~Hirschi, F.~Maltoni, O.~Mattelaer
  et~al., \emph{{The automated computation of tree-level and next-to-leading
  order differential cross sections, and their matching to parton shower
  simulations}}, \href{https://doi.org/10.1007/JHEP07(2014)079}{\emph{JHEP}
  {\bfseries 07} (2014) 079} [\href{https://arxiv.org/abs/1405.0301}{{\ttfamily
  1405.0301}}].
  
\bibitem{SMEFTMG}
C.~Degrande, G.~Durieux, F.~Maltoni, K.~Mimasu, A.~Vasquez, E.~Vryonidou, C.~Zhang. 
\href{https://feynrules.irmp.ucl.ac.be/wiki/SMEFTatNLO}{https://feynrules.irmp.ucl.ac.be/wiki/SMEFTatNLO}.

\bibitem{Sjostrand:2006za}
T.~Sjostrand, S.~Mrenna and P.~Z. Skands, \emph{{PYTHIA 6.4 Physics and
  Manual}}, \href{https://doi.org/10.1088/1126-6708/2006/05/026}{\emph{JHEP}
  {\bfseries 05} (2006) 026}
  [\href{https://arxiv.org/abs/hep-ph/0603175}{{\ttfamily hep-ph/0603175}}].

\bibitem{Sjostrand:2007gs}
T.~Sjostrand, S.~Mrenna and P.~Z. Skands, \emph{{A Brief Introduction to PYTHIA
  8.1}}, \href{https://doi.org/10.1016/j.cpc.2008.01.036}{\emph{Comput. Phys.
  Commun.} {\bfseries 178} (2007) 852}
  [\href{https://arxiv.org/abs/0710.3820}{{\ttfamily 0710.3820}}].

\bibitem{Conte:2012fm}
E.~Conte, B.~Fuks and G.~Serret, \emph{{MadAnalysis 5, A User-Friendly
  Framework for Collider Phenomenology}},
  \href{https://doi.org/10.1016/j.cpc.2012.09.009}{\emph{Comput. Phys. Commun.}
  {\bfseries 184} (2013) 222}
  [\href{https://arxiv.org/abs/1206.1599}{{\ttfamily 1206.1599}}].

\bibitem{Baur:1994ia}
U.~Baur, T.~Han and J.~Ohnemus, \emph{{Amplitude zeros in W+- Z production}},
  \href{https://doi.org/10.1103/PhysRevLett.72.3941}{\emph{Phys. Rev. Lett.}
  {\bfseries 72} (1994) 3941}
  [\href{https://arxiv.org/abs/hep-ph/9403248}{{\ttfamily hep-ph/9403248}}].

\bibitem{Dixon:1993xd}
L.~J. Dixon and Y.~Shadmi, \emph{{Testing gluon selfinteractions in three jet
  events at hadron colliders}},
  \href{https://doi.org/10.1016/0550-3213(94)90563-0}{\emph{Nucl. Phys. B}
  {\bfseries 423} (1994) 3}
  [\href{https://arxiv.org/abs/hep-ph/9312363}{{\ttfamily hep-ph/9312363}}].

\bibitem{CPo}
\emph{{Poisson distribution. Encyclopedia of Mathematics.}}
\href{https://encyclopediaofmath.org/index.php?title=Poisson_distribution}{https://encyclopediaofmath.org/index.php?title=Poisson{\_}distribution}

\bibitem{NEURIPS2019_9015}
A.~Paszke, S.~Gross, F.~Massa, A.~Lerer, J.~Bradbury, G.~Chanan et~al.,
  \emph{Pytorch: An imperative style, high-performance deep learning library},
  in \emph{Advances in Neural Information Processing Systems 32} (H.~Wallach, et al.), pp.~8024--8035.
\newblock Curran Associates, Inc., 2019.

\bibitem{deFavereau:2013fsa}
{\scshape DELPHES 3} collaboration, J.~de~Favereau, C.~Delaere, P.~Demin,
  A.~Giammanco, V.~Lema\^{i}tre, A.~Mertens et~al., \emph{{DELPHES 3, A modular
  framework for fast simulation of a generic collider experiment}},
  \href{https://doi.org/10.1007/JHEP02(2014)057}{\emph{JHEP} {\bfseries 02}
  (2014) 057} [\href{https://arxiv.org/abs/1307.6346}{{\ttfamily 1307.6346}}].

\end{thebibliography}

\providecommand{\href}[2]{#2}\begingroup\raggedright\endgroup

\end{document}